\documentclass[12pt]{article}
\usepackage{epsfig}
\usepackage{setspace}
\usepackage{amsbsy}
\usepackage{natbib}

\newcommand{\real}{\rm{I\!R}}
\newcommand{\T}{^{\rm T}}

\begin{document}
\title{Smooth blockwise iterative thresholding: \\ a smooth fixed point estimator based on\\ the likelihood's block gradient}
% for smooth model selection} %\\ against erratic Stein unbiased risk estimate}

%\runtitle{Smooth blockwise iterative thresholding}

\author{SYLVAIN SARDY\footnote{2-4 rue du Li\`evre, CP 64, 1211 Gen\`eve 4, Switzerland; sylvain.sardy@unige.ch}\\{\em Department of Mathematics, University of Geneva}}
\date{}

 \maketitle

\begin{quote}

The proposed smooth blockwise iterative thresholding estimator (SBITE) is a model selection technique defined as a fixed point reached by iterating
a likelihood gradient-based thresholding function.
The smooth James-Stein thresholding function has two regularization parameters $\lambda$ and $\nu$,
and a smoothness parameter $s$.
It enjoys smoothness like ridge regression and selects variables like lasso.
 %It regularizes the likelihood function and achieves model selection.
Focusing on Gaussian regression, we show that  SBITE is uniquely defined, and
that its Stein unbiased risk estimate  is a smooth function of $\lambda$ and $\nu$,
for better selection of the two regularization parameters.
%helps for the selection of the two regularization parameters for better predictive performance.
%In particular SBITE provides smooth versions of adaptive lasso and group lasso, of which we derive the smooth degrees of freedom
%to drive the selection of two regularization parameters. % are selected by minimizing a smooth and unbiased estimation of the risk.
% It also has additional flexibility thanks to more than one regularization parameter, and the ability to select
% these parameters well thanks to a smooth Stein unbiased risk estimate. 
%We revisit and propose smooth versions of adaptive-, group-, lasso, and derive their degrees of freedom.
%So far, only that of lasso is known.
We perform a Monte-Carlo simulation to investigate the predictive and oracle properties of this smooth version of adaptive lasso.

The motivation is a gravitational wave burst detection problem from several concomitant time series.
A nonparametric wavelet-based estimator is developed to combine information from all captors by block-thresholding multiresolution coefficients.
We study how the smoothness parameter $s$ tempers the erraticity  of the risk estimate,
 and derive a universal threshold, an information criterion and an oracle inequality in this canonical setting.

\end{quote}

%\begin{keyword}[class=AMS]
%\kwd[Primary ]{62G05, 62G08 }
%\kwd{62J07}
%\kwd[; secondary ]{62F10, 62J05, 62J12}
%\end{keyword}


%%%%%

%%%%%%%%%%%%%%%%%%%%%%
%%%%%%%%%%%%%%%%%%%%%% BEGINS HERE
\newpage

\section{Introduction}
\label{sct:intro}

% 
% \citet{Jame:Stei:esti:1961} showed the surprising result that for estimating a vector ${\boldsymbol \alpha} \in \real^P$ with $P>2$ from
% a Gaussian measurement ${\bf Y}\sim {\rm N}({\boldsymbol \alpha},I)$, the maximum likelihood estimate (MLE) is not admissible with respect to the $\ell_2$-risk.
% Indeed they showed that
% $
% \hat {\boldsymbol \alpha}^{{\rm JS}}({\bf Y})=(1-\frac{P-2}{\|{\bf Y} \|_2^2}) {\bf Y}
% $
% verifies
% $$
% {\rm R}(\hat {\boldsymbol \alpha}^{{\rm JS}},{\boldsymbol \alpha})<{\rm R}(\hat {\boldsymbol \alpha}^{{\rm MLE}},{\boldsymbol \alpha}) \quad \forall {\boldsymbol \alpha} \in \real^P,\ P>2,
% $$
% where ${\rm R}(\hat {\boldsymbol \alpha},{\boldsymbol \alpha})={\rm E}\|\hat {\boldsymbol \alpha}-{\boldsymbol \alpha} \|_2^2$.
% The James-Stein estimator above does not shrink (because the factor $1-\frac{P-2}{\|{\bf Y} \|_2^2}$ can be less than $-1$) and does not threshold (i.e., ${\rm P}(\hat {\boldsymbol \alpha}^{{\rm JS}}({\bf Y})={\bf 0})=0$). The drawback that the factor can be negative has been alleviated by taking its positive part
% \begin{equation}
% \hat {\boldsymbol \alpha}^{{\rm JS+}}({\bf Y})=(1-\frac{P-2}{\|{\bf Y} \|_2^2})_+ {\bf Y},
% \label{eq:JS+}
% \end{equation}
% which makes it not only a shrinkage estimator but also a thresholding estimator.
% Note that this latter estimator, although better than its previous version in terms of $\ell_2$-risk, remains inadmissible. 

Assuming a simple Gaussian model ${\bf Y}\sim {\rm N}({\boldsymbol \alpha},I)$ with ${\boldsymbol \alpha} \in \real^P$,
 the  \citet{Jame:Stei:esti:1961}  estimator 
%$\hat {\boldsymbol \alpha}^{{\rm JS}}=(1-\frac{P-2}{\|\hat {\boldsymbol \alpha}^{\rm MLE} \|_2^2}) \hat {\boldsymbol \alpha}^{\rm MLE}$
$\hat {\boldsymbol \alpha}^{{\rm JS}}= c \hat {\boldsymbol \alpha}^{\rm MLE}$ with $c=1-(P-2)/\|\hat {\boldsymbol \alpha}^{\rm MLE} \|_2^2$
proved that the maximum likelihood estimate $\hat {\boldsymbol \alpha}^{\rm MLE}$  is not admissible when $P>2$, since James Stein's mean squared error is
smaller for all coefficients.
This gave birth to a class of shrinkage or thresholding estimators of the form
\begin{equation}
\hat {\boldsymbol \alpha} = \eta_\lambda(\hat {\boldsymbol \alpha}^{\rm MLE}),
\label{eq:shrinkthresh}
\end{equation}
 where $\lambda$ controls the regularization. \emph{Shrinkage}  means that
$\| \hat {\boldsymbol \alpha}\| \leq \|\hat {\boldsymbol \alpha}^{\rm MLE} \|$, and
\emph{thresholding}  means that entries of $\hat {\boldsymbol \alpha}$ are set to zero to achieve variable selection.
When applied  coordinatewise to $\hat {\boldsymbol \alpha}^{\rm MLE}$, 
thresholding sets some entries of $\hat {\boldsymbol \alpha}$ to zero,
and when applied blockwise, then all entries are set to zero at once.
 The original James-Stein estimator  neither shrink nor threshold, but its truncated version
% $\hat {\boldsymbol \alpha}^{{\rm JS+}}=(1-\frac{P-2}{\|\hat {\boldsymbol \alpha}^{\rm MLE} \|_2^2})_+ \hat {\boldsymbol \alpha}^{\rm MLE}$
$\hat {\boldsymbol \alpha}^{{\rm JS+}}=c_+ \hat {\boldsymbol \alpha}^{\rm MLE}$
 does both blockwise by taking the positive part (i.e., $c_+=\max(c,0)$) of the multiplicative factor.
Waveshrink \citep{Dono94b} is a famous example of coordinatewise thresholding for wavelet smoothing.

Consider now generalized linear models \citep{NW72} with observed response $y_n$ and $P$ corresponding
covariates $\tilde {\boldsymbol x}_n=(\tilde x_{n,1}, \ldots, \tilde x_{n,P})$ organized in a matrix $\tilde X$ for $n=1,\ldots,N$,
negative log-likelihood $-l$ (including a possible link function) and coefficients ${\boldsymbol \alpha}$.
In the following, covariates have been mean-centered  and $\Sigma$-rescaled  
into a matrix $X$
%$X=(I-\frac{1}{N} J)\tilde X D_\Sigma$
 with corresponding coefficients ${\boldsymbol \beta}$,
so that $\hat {\boldsymbol \beta}^{\rm MLE}$ is homoscedastic in the rescaled basis \citep{SardyISI08}.
%to have unit entries on the diagonal of its covariance matrix.
%After estimating  $\hat {\boldsymbol \beta}$, %with a particular regularization technique,
% coefficients estimate $\hat {\boldsymbol \alpha}$ of the original problem are obtained as
%$\hat {\boldsymbol \alpha}=D_\Sigma \hat {\boldsymbol \beta}$ and
%$\hat \alpha_0=\bar {\bf Y}-\frac{1}{N} {\bf 1}^{\rm T} \tilde X\hat {\boldsymbol \alpha}$ for the intercept.
In that general regression setting, another class of regularization  defines the estimate
as a minimizer to a penalized likelihood function,
\begin{equation}
\hat {{\boldsymbol \beta}} = {\rm arg} \min_{{\boldsymbol \beta} \in {\cal B}} -l(X {\boldsymbol \beta}; {\bf y}) + \lambda \| {\boldsymbol \beta} \|,
\label{eq:penalizedlik}
\end{equation}
where  $\| \cdot \|$ is a norm or a semi-norm,
and $\lambda$ is the regularization parameter.
Famous examples of such methods are ridge regression \citep{ridgeHK}, %penalized likelihood density estimation \citep{GoodGaskins71},
nonparametric smoothing splines \citep{Wahb:spli:1990}, waveshrink, nonnegative garrote \citep{Brei:bett:1995} or lasso \citep{Tibs:regr:1996}.
The last three are variable selection estimators.
There exist exact links between (\ref{eq:shrinkthresh}) and (\ref{eq:penalizedlik}), for example waveshrink. % \citep{dono:john:hoch:ster:1992}.

One goal of this paper is to achieve variable selection with a new class of estimators called \emph{smooth blockwise iterative thresholding estimators (SBITE)},
that iteratively apply a new thresholding function called \emph{smooth James-Stein}, which is governed by a thresholding parameter $\lambda$,
a shrinkage parameter $\nu$ and a smoothness parameter $s$.
We will see this class of estimators can minimize penalized likelihood functions (\ref{eq:penalizedlik}) by iteratively applying smooth James-Stein thresholding
for $(\lambda,\nu,s)$ set to specific values.
%For instance, lasso and soft-Waveshrink with $s=\nu=1$, and hard-Waveshrink with $s=1$ and letting $\nu \rightarrow \infty$.
In that sense iterative thresholding encompasses existing variable selection methods of the type (\ref{eq:shrinkthresh}) and (\ref{eq:penalizedlik}) as particular cases.
Iterative thresholding goes beyond these methods by adding flexibility, stability, smoothness and uniqueness properties.
Iterative thresholding can  be done coordinatewise or blockwise.
As far as flexibility is concerned, recent estimators are governed by several regularization parameters, for example, bridge \citep{Fu:1998}, SCAD \citep{Fan:Li:vari:2001},
the penalized least squares estimator of  \citet{AntoFan01}, EBayesThresh \citep{JS04,JS05},
fused lasso \citep{Tibs:Saun:Ross:Zhu:Knig:spar:2005}, elastic net \citep{ZouHastie05}, adaptive lasso \citep{Zou:adap:2006} and $\ell_\nu$-regularization \citep{SardySLIC09}.
It is often believed that, although these estimators are more flexible, their risk may be worse in some  situations
because selection of more than one regularization parameter is unstable. 
A smooth estimation of the risk with SBITE will allow a more stable selection of the regularization parameters.
SBITE also owes its stability  \citep{B96} to its smoothness like ridge regression.
As far as uniqueness and smoothness are concerned, we will see that smooth James-Stein iterative thresholding smoothly and uniquely extends lasso, which otherwise
is not smooth and not necessarily uniquely defined.

%One goal of this paper is to prove the contrary thanks to a new thresholding function called smooth James-Stein.

% We consider two situations where we employ smooth James-Stein thresholding.
This paper is organized as follows.
Section~\ref{sct:motiv} presents our original motivation, the detection of gravitational wave bursts using information from several simultaneously recorded time series.
It motivates the need for a wavelet-based smoother that thresholds blocks of multiresolution coefficients across captors.
Section~\ref{sct:SBITE} presents the new SBIT estimator for generalized linear regression,
discusses its links to existing estimators, presents uniqueness and smoothness properties,
and derives its Stein unbiased risk estimate. % that has the advantage of being smooth to help better select the regularization parameters.
A Monte-Carlo experiment investigates its finite sample properties. % in comparison with existing estimators.
Section~\ref{sct:bcp} focuses on  block canonical regression, %(i.e., Gaussian regression with identity matrix with coefficients organized by block).
where we study tempering of the erratic behavior of the Stein unbiased risk estimate
by means of the smoothness coefficient of SBITE, and derive a universal threshold, an information criterion and an oracle inequality.
Finally the estimator is applied to gravitational wave bursts
and simulated data. % in various settings parametrized by level of sparsity, signal-to-noise ratio and number of captors.
Section~\ref{sct:further} discusses two extensions.

%The corresponding estimator is indexed by two hyperparameters for more flexibility and a smoothness parameter for better estimation of its $\ell_2$-risk with the Stein unbiased risk estimate (SURE).

\section{Motivation}
\label{sct:motiv}

Gravitational wave bursts are produced by energetic cosmic phenomena such as the collapse of a supernova.
They are  rare, highly oscillating and of small intensity compared to the instrumental noise,
so only the concomitant recording by $Q$ captors (typically $Q=3$) of the cosmic phenomena may help prove the existence of such wave bursts.
The captors are located far apart from each other to avoid recording local earth phenomena (such as an earthquake) on all $Q$ captors.
Another difficulty is the non-white nature of the noise, possibly non-Gaussian. The measurements are recorded at a high frequency 
of $5$ KHz: one minute of recording has $3Q\cdot 10^5$ noisy measurements. A good model for these data is
\begin{equation}
\tilde S_{t}^{(q)}=\mu^{(q)}(t)+\tilde \epsilon_{t}^{(q)}, \quad t=1,\ldots,T, \ q=1\ldots,Q
\label{eq:modelgw}
\end{equation}
where the noises $\tilde \epsilon^{(q)}$ and $\tilde \epsilon^{(q')}$ are independent between captors $q \neq  q'$,
but where the noise is temporally correlated for a given captor.
Importantly, most of the time  the underlying signal $\mu^{(q)}(t)=0$ for all $q$, and, if $\mu^{(q)}(t) \neq 0$
for a given time $t$ and captor $q$, then the same is true for all the other captors.
Finally because the incoming wave burst may not hit the captors with the same angle,
we may not have $\mu^{(q)}(t)=\mu^{(q')}(t)$, but only a proportionality constant relates them. 

Like  \citet{gravitwave04}, we assume each underlying signal $\mu^{(q)}$ expands on $T=N$ orthonormal approximation
$\phi$ and fine scale $\psi$ wavelets:
\begin{equation}
\mu^{(q)}(t)=\sum_{k=0}^{2^{j_0}-1} {\beta_0}_{k}^{(q)} \phi_{j_0,k}(t) +
\sum_{j=j_0}^J\sum_{n=0}^{N_j-1} {\beta}_{j,n}^{(q)} \psi_{j,n}(t),
\label{eq:wavelet-based-expansion}
\end{equation}
where the wavelets are obtained by dilation $j$ and translation $n$, $J=\log_2(N)$ and $N_j=2^j$; see \citet{Dono94b}.
One can extract an orthonormal regression matrix $W=[\Phi_0 \Psi_{j_0} \dots \Psi_{J}]$  from this representation such that (\ref{eq:modelgw}) becomes $\tilde {\boldsymbol S}^{(q)} = W {\boldsymbol \beta}^{(q)}+
\tilde { \boldsymbol \varepsilon}^{(q)}$. Applying the orthonormal wavelet decomposition $W\T$, the model can also be written as
%\begin{equation}
 ${\boldsymbol S}^{(q)}={\boldsymbol \beta}^{(q)}+{ \boldsymbol \varepsilon}^{(q)}$,
%\label{eq:Wt}
%\end{equation}
where ${\boldsymbol S}^{(q)}=W\T \tilde {\boldsymbol S}^{(q)}$ and ${ \boldsymbol \varepsilon}^{(q)}=W\T \tilde { \boldsymbol \varepsilon}^{(q)}$.
\begin{figure}[!ht]
  \begin{center}
% \centerline{\includegraphics[height=6in,width=3in]{paperfig1}}
\includegraphics[height=3.0cm, angle=-90,width=12cm]{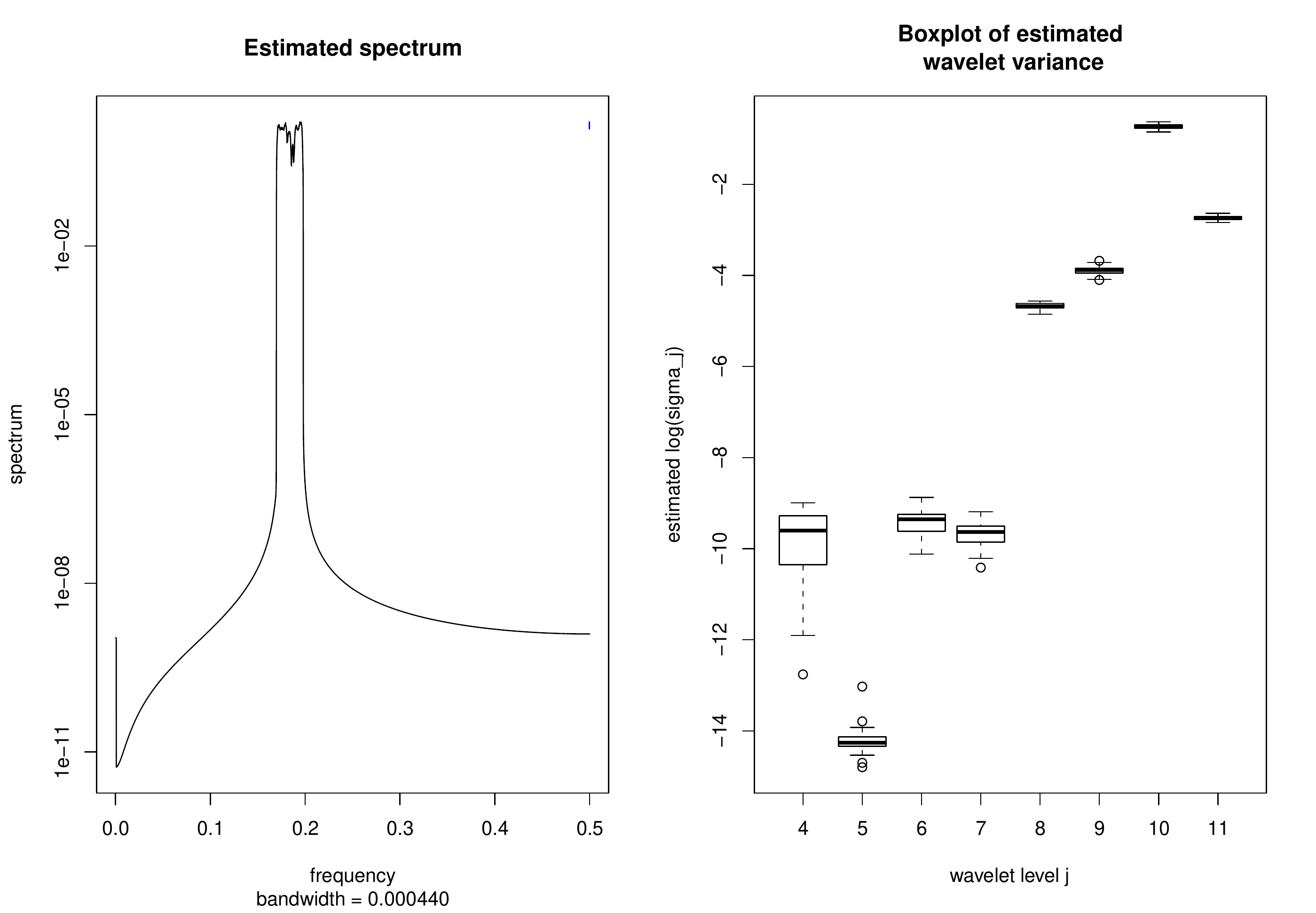}
  \caption{EDA from 26 seconds recording: (left) estimated spectrum and (right) boxplots of estimated wavelet variances on a log-scale
for level 4 to 11.}
  \label{fig:spectrandwavevar}
  \end{center}
\end{figure}
This latter model is interesting in three aspects.
First \citet{John:Silv:wave:1997} show that stationary correlated noise $\tilde { \boldsymbol \varepsilon}_q$ is well decorrelated by a wavelet transform within and between levels.
And for a given level $j$, the nearly white noise process has its own wavelet variance $\sigma_{j}^{2,(q)}$ \citep{Perc:on:1995,Serr:Wald:Perc:stat:2000}
 that can be estimated from the data \citep{Dono95i}.
The right plot of Figure~\ref{fig:spectrandwavevar} represents boxplots of such wavelet variances
estimated from $31$ disjoint signals of length $N=4096$ for $j=4,\ldots,11$.
The left plot of Figure~\ref{fig:spectrandwavevar} shows an estimated spectrum from 26 seconds of recording: the noise of the captor
is a band-pass filtered colored noise process.
Second, the data are Gaussianized by the linear wavelet transformation.
Third, owing to sparse wavelet representation, the vectors ${\boldsymbol \beta}^{(q)}$ are sparse,
and the amount of sparsity varies between levels, so the selection of hyperparameters should be level dependent.
Hence organizing the coefficients %${\boldsymbol \beta}^{(q)}=({\boldsymbol \beta}^{(q)}_{j_0}, \ldots, {\boldsymbol \beta}^{(q)}_{J})$
 of captor $q$ by levels,
model (\ref{eq:modelgw}) and (\ref{eq:wavelet-based-expansion}) can be well approximated at a given level $j$ by
\begin{equation}
 {\boldsymbol Y}_{j}^{(q)}={\boldsymbol \alpha}_{j}^{(q)}+{ \boldsymbol z}_{j}^{(q)} \quad {\rm with} \quad
{\boldsymbol Y}_{j}^{(q)}={\boldsymbol S}_{j}^{(q)}/\sigma_{j}^{(q)}  \quad {\rm and} \quad
          {\boldsymbol \alpha}_{j}^{(q)}={\boldsymbol \beta}_{j}^{(q)}/\sigma_{j}^{(q)},
%\left \{ \begin{array}{c}
%          {\boldsymbol Y}_{j}^{(q)}={\boldsymbol S}_{j}^{(q)}/\sigma_{j}^{(q)} \\
%          {\boldsymbol \alpha}_{j}^{(q)}={\boldsymbol \beta}_{j}^{(q)}/\sigma_{j}^{(q)} 
%         \end{array} \right . ,
\label{eq:Y_j}
\end{equation}
where ${\boldsymbol \alpha}_{j}^{(q)}=(\alpha_{j,1}^{(q)},  \ldots, \alpha_{j,N_j}^{(q)})$ is a sparse vector  and ${ \boldsymbol z}_{j}^{(q)} \stackrel{{\rm i.i.d.}}
\sim {\rm N}(0,1)$.
Importantly, given a dilation $j$ and a translation $n$, then
$
{\boldsymbol \alpha}_{j,n}=(\alpha_{j,n}^{(1)}, \ldots, \alpha_{j,n}^{(Q)})
$
is either null %${\boldsymbol \alpha}_{j,n}={\boldsymbol 0}$
%(due to no wave burst, or to no component of the wave burst on wavelet $n$ at level $j$),
or, when ${\boldsymbol \alpha}_{j,n}\neq {\bf 0}$, then its entries  are  different.
% (due to captors facing the arriving wave burst at different angles, or to different wavelet variances $\sigma_{j}^{2,(q)}$ and $\sigma_{j}^{2,(q')}$ for captors $q\neq q'$).
Our goal is to derive an estimator that adapts levelwise to block sparsity.

\section{Smooth blockwise iterative thresholding}
\label{sct:SBITE}

\subsection{Review of block coordinate relaxation}
\label{subsct:BCR}

Recent estimators consider the situation where the coefficients are blocked into $J$ groups
${\boldsymbol \beta}=({\boldsymbol \beta}_1,\ldots,{\boldsymbol \beta}_J)$ of respective sizes $p_1, \ldots, p_J$ with $\sum_{j=1}^J p_j=P$.
Correspondingly, let $X=[X_1 \ldots X_J]$.
For instance, for gravitational wave burst detection,  wavelet coefficients are grouped into blocks of  size $Q$, the number of captors,
and $X_j$ is the $Q\times Q$ identity matrix for all $j=1,\ldots,J$.

We recall an optimization technique upon which we elaborate a new estimator in the following section. Suppose for now we want to calculate
$\hat {\boldsymbol \beta}^{\rm MLE}$ solution to (\ref{eq:penalizedlik}) for $\lambda=0$.
%$
%\hat {\boldsymbol \beta}^{\rm MLE}= {\rm arg} \min_{{\boldsymbol \beta} \in {\cal B}} -l(X {\boldsymbol \beta}; {\bf y}).
%$ 
Block coordinate relaxation (BCR)  works as follows:
start with any initial guess ${\boldsymbol \beta}$, choose a block $j\in\{1,\ldots,J\}$ and update only the $j$th block ${\boldsymbol \beta}_j$
conditional on the values of the other blocks ${\boldsymbol \beta}_i$ for all $i \neq j$, that is
\begin{equation}
{\boldsymbol \beta}_j^{\mid {\rm MLE}}= {\rm arg} \min_{{\boldsymbol \beta}_j \in {\cal B}_j} -l(\sum_{i\neq j}X_i {\boldsymbol \beta}_i + X_j {\boldsymbol \beta}_j ; {\bf y}),
\label{eq:update}
\end{equation}
and leave the other unchanged to obtain the next iterate
\begin{eqnarray}
{\boldsymbol \beta}^{(j)}=({\boldsymbol \beta}_1, \ldots, {\boldsymbol \beta}_{j-1}, &{\boldsymbol \beta}_j^{\mid {\rm MLE}}&, {\boldsymbol \beta}_{j+1}, \ldots, {\boldsymbol \beta}_J).
\label{eq:betaupdate} %\\
%{\boldsymbol \beta}^{(j)\rightarrow 0}=({\boldsymbol \beta}_1, \ldots, {\boldsymbol \beta}_{j-1}, &{\bf 0}&, {\boldsymbol \beta}_{j+1}, \ldots, {\boldsymbol \beta}_J) \label{eq:betaupdate0} 
\end{eqnarray}
Note that $j$ is the index of the updated block, but not of the iteration.

\emph{Property~1}: Assuming that ${\boldsymbol \beta}  \in {\cal B}$, where ${\cal B}={\cal B}_1\times \ldots \times {\cal B}_J$ is a product
of closed convex sets, 
that (\ref{eq:update}) has a unique solution and that the log-likelihood is continuously differentiable, then the algorithm converges to a stationary point \citep[Proposition~2.7.1]{Bert}.
If the negative log-likelihood is also strictly convex, then the algorithm finds the MLE.

\emph{Property~2}: After updating ${\boldsymbol \beta}$ with ${\boldsymbol \beta}^{(j)}$ according to~(\ref{eq:betaupdate}),
the gradient of the likelihood with respect to the $j$th block is null, that is
$\nabla_{{\boldsymbol \beta}_j} l(X {\boldsymbol \beta}^{(j)} ; {\bf y}) ={\bf 0}$.

%%%%%%%%%%%%%%
\subsection{Smooth blockwise iterative thresholding estimator}

The MLE does not achieve variable selection however. To do so, one can test  the significance of the $j$th block 
based on the likelihood and its gradient in the following way.
Suppose we are at the MLE where the entire gradient vector is null.
The covariates have been $\Sigma$-rescaled for all MLE coefficients
to have unit variance as discussed in Section~\ref{sct:intro}.
So if after thresholding the $j$th block of the MLE to zero the gradient with respect to the $j$th block remains small (compared to a threshold $\lambda$),
 then we declare this block not significant.
This suggests calculating the likelihood's block gradient at each BCR iteration:
  \begin{itemize}
    \item at the current iterate ${\boldsymbol \beta}^{(j)}$ defined  by (\ref{eq:betaupdate}). According to  Property~2, the gradient with respect to the $j$th block
is null;
    \item at the current iterate with ${\boldsymbol \beta}_j^{\mid {\rm MLE}}$ thresholded to ${\bf 0}$, namely at
\begin{eqnarray}
%{\boldsymbol \beta}^{(j)}=({\boldsymbol \beta}_1, \ldots, {\boldsymbol \beta}_{j-1}, &{\boldsymbol \beta}_j^{\mid {\rm MLE}}&, {\boldsymbol \beta}_{j+1}, \ldots, {\boldsymbol \beta}_J)
%\label{eq:betaupdate} \\
{\boldsymbol \beta}^{(j)\rightarrow 0}=({\boldsymbol \beta}_1, \ldots, {\boldsymbol \beta}_{j-1}, &{\bf 0}&, {\boldsymbol \beta}_{j+1}, \ldots, {\boldsymbol \beta}_J).
\label{eq:betaupdate0} 
\end{eqnarray}
The gradient  with respect to the $j$th block is
$\nabla_{{\boldsymbol \beta}_j} l(X {\boldsymbol \beta}^{(j)\rightarrow 0} ; {\bf y})$.
  \end{itemize}
A difference larger than a threshold $\lambda$ between the two likelihood's block gradient norms, that is,
$
\|\nabla_{{\boldsymbol \beta}_j} l(X {\boldsymbol \beta}^{(j)\rightarrow 0} ; {\bf y})\| \geq \lambda
$, shows that the $j$th block is significant given the value of the other blocks. This leads to the following estimator.

\emph{Smooth block iterative thresholding estimator (SBITE)} (algorithmic definition). Choose a threshold $\lambda \geq 0$,
a shrinkage parameter $\nu\geq 1$ and a smoothness parameter $s\geq1$.
Let $\tilde {\boldsymbol \beta}^* $  be a root-$N$-consistent estimate of ${\boldsymbol \beta}$.
\begin{enumerate}
 \item Start with any initial value;
 \item Choose a block $j$, and calculate ${\boldsymbol \beta}_j^{\mid {\rm MLE}}$ according to (\ref{eq:update}) and the gradient $\nabla_{{\boldsymbol \beta}_j} l(X {\boldsymbol \beta}^{(j)\rightarrow 0} ; {\bf y})$;
 \item Update the $j$th block according to
\begin{equation}
  {\boldsymbol \beta}_j^{\rm update}=(1-\frac{\lambda^\nu}{\|\tilde{\boldsymbol \beta}_j^* \|^{\nu-1} \|\nabla_{{\boldsymbol \beta}_j} l(X {\boldsymbol \beta}^{(j)\rightarrow 0} ; {\bf y}) \|})^s_+
  {\boldsymbol \beta}_j^{\mid {\rm MLE}};
\label{eq:smooththreshold}
\end{equation}
 \item Go back to step 2 until convergence.
\end{enumerate}
The thresholding function (\ref{eq:smooththreshold}) is called \emph{smooth James-Stein}:
when $\|\tilde{\boldsymbol \beta}_j^* \|^{\nu-1}$ and $\|\nabla_{{\boldsymbol \beta}_j} l(X {\boldsymbol \beta}^{(j)\rightarrow 0} ; {\bf y}) \|$ are small,
 thresholding sets the $j$th block to zero.
We study the advantage of the smoothness parameter $s$ later.
%Observe in (\ref{eq:smooththreshold}) that not only a larger gradient but also a larger root-$N$-consistent estimate $\hat {\boldsymbol \beta}_j$
%of the $j$th block in (\ref{eq:smooththreshold}) will lead to milder shrinkage/thresholding for a given $\lambda$.
For $s=1$ and certain values of $\nu$,
SBITE is linked to existing estimators,  as established in the following property.

\emph{Property~3} (Gaussian case with a smoothness parameter $s=1$): The SBITE iterations converge  at the limit to the estimate of:
\begin{enumerate}
 \item lasso \citep{Tibs:regr:1996} for $s=\nu=1$ and blocks of size one. Lasso is not an oracle procedure \citep{FanLi01}.
 \item group lasso \citep{bakin99,Yuan:Lin:mode:2006} for $s=\nu=1$, with blocks.
 \item adaptive lasso, which is oracle \citep{Zou:adap:2006}, for $s=1$ and $\nu> 1$ and blocks of size one. Adaptive lasso is the motivation
for including the norm $\|\tilde{\boldsymbol \beta}_j^* \|$  in (\ref{eq:smooththreshold}). Hence
not only a larger gradient but also a larger root-$N$-consistent estimate
of the $j$th block  leads to milder shrinkage.
 \item waveshrink for $s=1$ and $X$ an orthonormal wavelet matrix: soft-waveshrink for $\nu=1$  and  hard-waveshrink when $\nu \rightarrow \infty$.
 \item truncated James-Stein for $\nu=2$,  $\lambda=\sqrt{P-2}$ and a block of size $P>2$.
 \item block thresholding for wavelet smoothing for $s=1$, $\nu=2$, groups of size $L>1$ and $X$ an orthonormal wavelet matrix \citep{Cai:adap:1999}.
% \item smooth adaptive group lasso for $s\geq1$, $\nu\geq 1$ and blocks of various sizes.
\end{enumerate}

For a proof, observe that the SBITE algorithm corresponds to the shooting algorithm \citep{Fu:1998} for lasso, the BCR algorithm \citep{sardyJCGS}
for basis pursuit \citep{CDS99}
 and to the iterative algorithm of \cite{Yuan:Lin:mode:2006} to minimize penalized least squares problems of the form
%We recall that the adaptive group lasso estimate is solution to
\begin{equation}
\min_{ {\boldsymbol \beta}} \frac{1}{2} \|{\bf Y}-X {\boldsymbol \beta} \|_2^2 + \lambda \sum_{j=1}^J \frac{1}{\|\tilde{\boldsymbol \beta}_j^*\|^{\nu-1}}\|{\boldsymbol \beta}_j\|_2.
\label{eq:adaptlasso}
\end{equation}
%where $\hat{\boldsymbol \beta}_j$ is a root-$N$-consistent estimate of ${\boldsymbol \beta}_j$.
Note that the last three estimators above (4, 5, 6) converge after one iteration, and (\ref{eq:adaptlasso}) can also be solved by another class of iterative algorithms developed for inverse problems \citep{DaubInvProb04}.
% 
% Convergence of the SBITE algorithm  is known when $s=1$ in the Gaussian case \citep{Fu:1998,sardyJCGS}.
% Next section proves uniqueness of such a fixed point when $s>1$, but convergence to the fixed point remains to be proved.

Hence SBITE provides a new interpretation of lasso as a sequence of tests based on the block gradient of the likelihood evaluated at successive null hypothesis $H_0: {\boldsymbol \beta}_j={\bf 0}$ given the values of the other coefficients, until an equilibrium is reached.
A legitimate question addressed in the following section is whether such an equilibrium can be reached at a unique point. If so, SBITE is defined uniquely.

\subsection{Uniqueness}

Lasso ($s=1$) does not necessarily define a unique estimate if the kernel of the regression matrix $X$ is not the ${\bf 0}$ singleton \citep{SardySLIC09}.
On the contrary SBITE is uniquely defined under a milder condition when $s>1$, as stated in Theorem~1 below.
We first give its fixed point definition.

\emph{Smooth block iterative thresholding estimator (SBITE)} (fixed point definition). Choose a threshold $\lambda \geq 0$,
a shrinkage parameter $\nu\geq 1$ and a smoothness parameter $s\geq1$.
Let $\tilde {\boldsymbol \beta}^*$  be a root-$N$-consistent estimate of ${\boldsymbol \beta}$. SBITE is a fixed point to the SBITE algorithm,
which is the solution to the set of $P$ nonlinear equations
\begin{equation}
  {\boldsymbol \beta}_j=(1-\frac{\lambda^\nu}{\|\tilde{\boldsymbol \beta}_j^* \|^{\nu-1} \|\nabla_{{\boldsymbol \beta}_j} l(X {\boldsymbol \beta}^{(j)\rightarrow 0} ; {\bf y}) \|})^s_+
  {\boldsymbol \beta}_j^{\mid {\rm MLE}}, \quad j=1,\ldots,J,
\label{eq:fixedpointdef}
\end{equation}
 where ${\boldsymbol \beta}_j^{\mid {\rm MLE}}$ is defined by (\ref{eq:update}), and each ${\boldsymbol \beta}_j$ is a vector of length $p_j$ with $P=\sum_{j=1}^J p_j$.

\bigskip
Despite being highly non-linear and employing
a non-convex thresholding function, these equations define SBITE uniquely  in the Gaussian case when  $s>1$.

\bigskip
\emph{Theorem~1}: For the Gaussian likelihood, the solution to (\ref{eq:fixedpointdef}) with smoothness parameter $s>1$ 
 is uniquely defined for all matrices $X=[X_1 \ldots X_J]$ such that $X_j^{\rm T} X_j$ are positive definite matrices for all $j=1,\ldots,J$.
It is moreover continuously differentiable with respect to the data.

\bigskip
Note that the condition is milder than $X^{\rm T}X$ being positive definite;
for unit block size, its means that each column of $X$ must be different from the zero-vector.
Convergence of the SBITE algorithm is proved when $s=1$ for the Gaussian likelihood \citep{Fu:1998,sardyJCGS}
and for more general likelihoods \citep{SardyTseng04}.
Convergence to the unique fixed point has always been observed when $s>1$, but remains to be proved.

\subsection{Equivalent degrees of freedom}
\label{subsct:edf}

SBITE is governed by two regularization parameters $\lambda$ and $\nu$, and a smoothness parameter $s$. %, used in the definition of the smooth James-Stein thresholding function (\ref{eq:smooththreshold}).
%To select them, consider the (Section~\ref{sct:further} considers the group case),
%In practice $\sigma$ can be estimated and the responses divided by it to approximate the assumption.
For $s=1$ and for the Gaussian linear model with mean ${\boldsymbol \mu} = X {\boldsymbol \alpha}$ and
$P$ variables grouped into blocks of unit size, \citet{Zou:adap:2006} selects the two regularization parameters $\lambda$ and $\nu$ of adaptive lasso by cross-validation,
a rule known for its high computational cost and instability.
This section derives instead the Stein unbiased risk estimate for any combination of the three parameters $(\lambda,\nu,s)$.
%which moreover has the property to be smoother the larger $s$.
%Importantly when $s>1$, the derived equivalent degrees of freedom has the advantage of being smooth.

\cite{Stein:1981} showed that for  an estimator of the form $\hat {\boldsymbol \mu}= g({\bf Y})+{\bf Y}$ and unit variance,
then the quadratic risk can be estimated unbiasedly if $g$ is almost differentiable.
For SBITE  with blocks of unit size, we have $g({\bf Y};\lambda,\nu,s)=\bar {\bf Y} {\bf 1}+X \hat {\boldsymbol \beta}_{\lambda,\nu;s}({\bf Y})-{\bf Y}$,
where $\hat {\boldsymbol \beta}_{\lambda,\nu;s}({\bf Y})$ is the solution to (\ref{eq:fixedpointdef}) for Gaussian likelihood.
SBITE is almost differentiable for $s=1$ and differentiable for $s>1$, so the Stein unbiased risk estimate (SURE) for SBITE is
%So we can employ the Stein unbiased risk estimate to select the hyperparameters $(\lambda,\nu)$ of SBITE by minimizing
\begin{equation}
{\rm SURE}(\lambda,\nu;s)={\rm RSS}(\hat {\boldsymbol \mu}_{\lambda,\nu,s})+ N +2 \sum_{n=1}^N \partial g_n({\bf Y};\lambda,\nu,s)/\partial Y_n,
\label{eq:SUREgeneral}
\end{equation}
where the last term is the so-called equivalent degrees-of-freedom.
SURE involves the partial derivatives 
%\begin{equation}
$\partial g_n({\bf Y};\lambda,\nu,s)/\partial Y_n =1/N+{\bf x}_{n}^{\rm row}\cdot \nabla_n \hat {\boldsymbol \beta}_{\lambda,\nu,s}({\bf Y})-1$, for $n=1,\ldots,N$,
%\label{eq:partialg}
% \end{equation}
where $\nabla_n \hat {\boldsymbol \beta}_{\lambda,\nu,s}({\bf Y})$ are the derivatives of $ \hat {\boldsymbol \beta}_{\lambda,\nu,s}({\bf Y})$ with respect to $Y_n$,
and ${\bf x}_{n}^{\rm row}$ is the $n$th row of $X$.
The following theorem states that $\nabla_n \hat {\boldsymbol \beta}_{\lambda,\nu,s}({\bf Y})$
is explicitly defined as solution to a full rank system of linear equations when $s>1$.
Hence the risk of SBITE can be estimated unbiasedly for all regression matrix.

\bigskip
\emph{Theorem~2}:
For a given pair~$\lambda>0$ and $\nu\geq 1$
and for  a \emph{linear} estimate $\tilde {\boldsymbol \beta}^*=A {\bf Y}$ (e.g., least squares or ridge regression)
where the entries of $A$ are noted $a_{pn}$, % for $n=1,\ldots,N$ and $p=1,\ldots,P$,
the gradient of the SBITE estimate $\hat {\boldsymbol \beta}_{\lambda,\nu,s}({\bf Y})$ with respect to $Y_n$ is the solution
to a system of linear equations (\ref{eq:lineqgrad}) that is full rank when $s>1$, regardless of the existence of a kernel for $X$,
for $n=1,\ldots,N$.

\bigskip

%and 
%\frac{\eta^{\rm soft}_{\lambda^\nu}(|\hat \alpha_p^{\rm OLS}|^{\nu-1}{\bf r}_{-p}^{\rm T} {\bf x}_p)}{s  \hat \alpha_p({\bf Y}) |\hat \alpha_p^{\rm OLS}|^{\nu-1} \|{\bf x}_p \|_2^2}$
%with
% \begin{eqnarray}
% u_p&=&\frac{s (\nu-1)a_{pn}(w_p-1){\bf r}_{-p}^{\rm T} {\bf x}_p}{\hat \alpha_p^{\rm LS}w_p^s},  \nonumber \\
% % v_p&=&\frac{s (\nu-1){\rm sign}(\hat \alpha_p^{\rm LS}) {\rm sign}({\bf r}_{-p}^{\rm T} {\bf x}_p) a_{pn} e_p^{s-1}}{|\hat \alpha_p^{\rm LS}|^{s\nu-s} |{\bf r}_{-p}^{\rm T} {\bf x}_p|^{s-1}}
% % \{ |\hat \alpha_p^{\rm LS}|^{\nu-2} |{\bf r}_{-p}^{\rm T} {\bf x}_p| - e_p/ |\hat \alpha_p^{\rm LS}|
% % \}, \nonumber \\
% % u_p&=&\frac{e_p^{s-1}}{|\hat \alpha_p^{\rm LS}|^{s\nu-s} |{\bf r}_{-p}^{\rm T} {\bf x}_p|^s}
% % \{ (1-s)e_p+
% % s |\hat \alpha_p^{\rm LS}|^{\nu-1} |{\bf r}_{-p}^{\rm T} {\bf x}_p| \}, \nonumber \\
% w_p &=& \frac{|\hat \alpha_p^{\rm LS}|^{\nu-1}|{\bf r}_{-p}^{\rm T} {\bf x}_p|}{e_p},  \label{eq:uve} \\
% e_p&=&\eta^{\rm soft}_{\lambda^\nu}(|\hat \alpha_p^{\rm LS}|^{\nu-1}|{\bf r}_{-p}^{\rm T} {\bf x}_p|). \nonumber
% \end{eqnarray}
% \begin{equation}
% 1/w_p=1-\frac{\lambda^\nu}{|\hat \beta_p|^{\nu-1}| r_{p}|} \quad {\rm and} \quad
% u_p=\frac{s (\nu-1)a_{pn}(w_p-1)r_{p}}{\hat \beta_p w_p^s}. \label{eq:uve} 
% \end{equation}
 %for  $p=p_i \in \bar {\cal I}_0$ and $i=1,\ldots,{|{\cal I}_0|}$.

Interestingly also from Theorem~2,
letting the smoothness parameter $s$ tend to one leads to the equivalent degrees of freedom
of adaptive lasso; if moreover $\nu=1$, then the solution to %the system of linear equations
  (\ref{eq:lineqgrad}) is  
$
 {\bf h}_n^{\bar {\cal I}_0}=((X^{\bar {\cal I}_0})\T X^{\bar {\cal I}_0})^{-1}( {\bf x}_n^{\rm row})\T
$.
Hence, we see that
 \begin{eqnarray*}
\sum_{n=1}^N \partial g_n({\bf Y}; \lambda, \nu, s)/\partial Y_n %&=&-N +1+ \sum_{n=1}^N  {\bf x}_n^{\rm row} \cdot \nabla_n \hat \beta({\bf Y}) \\
&=& 1+ \sum_{n=1}^N  {\bf x}_n^{\rm row} \cdot ((X^{\bar {\cal I}_0})\T X^{\bar {\cal I}_0})^{-1}({\bf x}_n^{\rm row})\T -N \\
&=& 1+ {\rm trace} (((X^{\bar {\cal I}_0})\T X^{\bar {\cal I}_0})^{-1} ((X^{\bar {\cal I}_0})\T X^{\bar {\cal I}_0})) -N \\
&=& 1+ |\bar {\cal I}_0| -N ,
\end{eqnarray*}
where $|\bar {\cal I}_0|$ is lasso's degrees of freedom previously found by \citet{ZHT07}.

\begin{figure}[!ht]
  \begin{center}
% \centerline{\includegraphics[height=6in,width=3in]{paperfig1}}
\includegraphics[angle=-90,width=13cm]{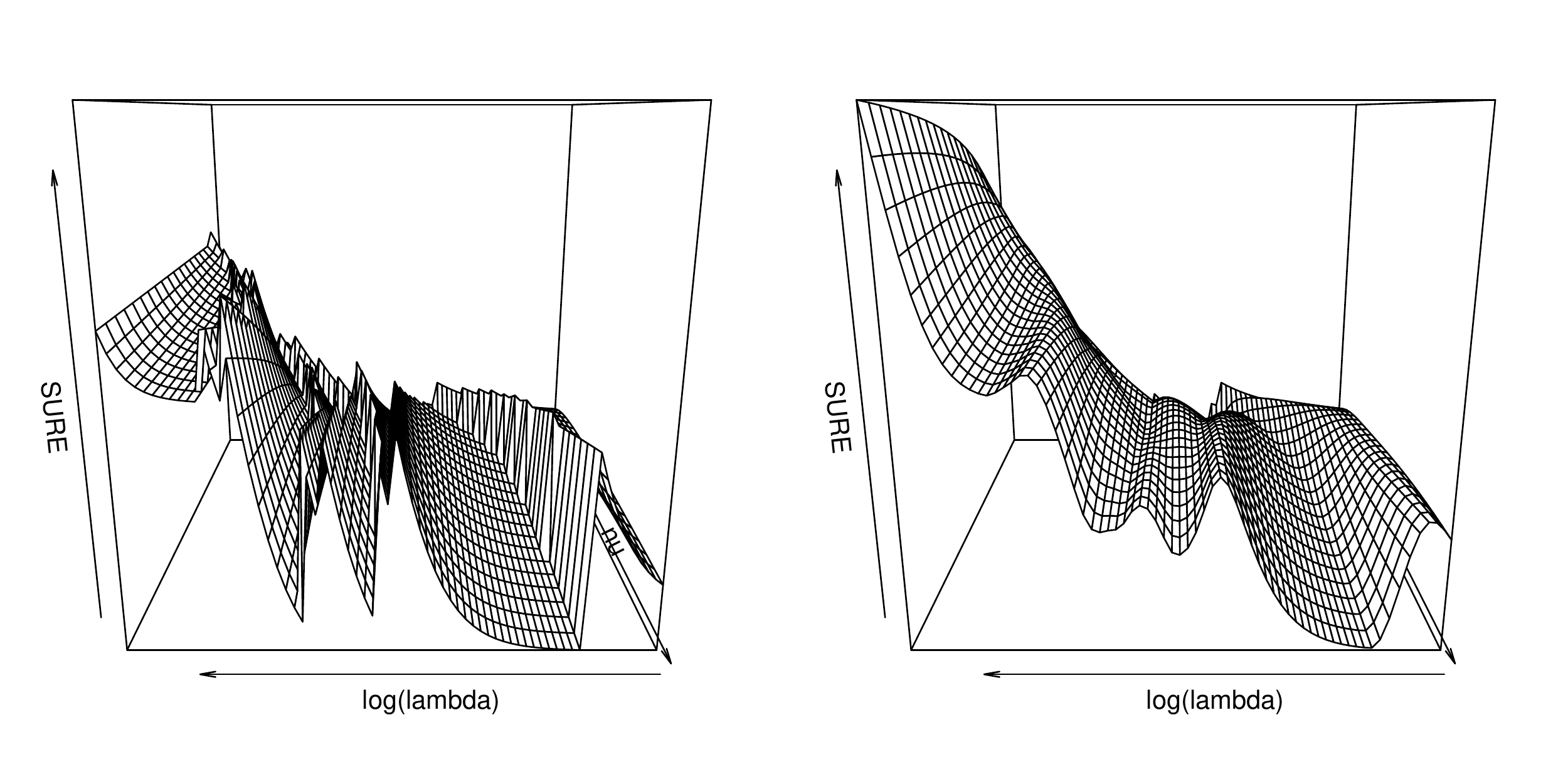}
  \caption{Prostate cancer data: Stein unbiased risk estimate as a function of $\lambda$ and $\nu$ for adaptive lasso (left, $s=1$)
and smooth adaptive lasso (right, $s=2\log \nu+1$).}
  \label{fig:prostate}
  \end{center}
\end{figure}

% 
%  a new result and a simple proof of an existing result.
% First, Theorem~2 leads to the equivalent degrees of freedom
% of adaptive lasso ($s=1$) letting $D_{p,p}^{\bar {\cal I}_0}=1$  in (\ref{eq:Dpii}) for all $p$.
% Second, the lasso (i.e., $s=1$ and $\nu=1$) has $u_{p}=0$ in (\ref{eq:zni}) and (\ref{eq:lineqgrad}), so that 
% \bigskip
The smoothness parameter should not be considered as a third regularization parameter like $\lambda$ and $\nu$, but more like a device
to bring smoothness to the estimator and combat the increasing erraticity of the two-dimensional SURE function  as $\nu$ grows. 
Section~\ref{subsct:TVSURE} quantifies SURE's erraticity tempered with the smoothness parameter~$s$.
The smoothness parameter  should not be too large however, since it contradicts the goal of
 a large $\nu$ to approach hard thresholding, and since the constant of the oracle inequality (\ref{eq:OrIneq}) of Theorem~4 increases with $s$. 
A good trade-off is for instance $s(\nu)=2\log \nu +1$ (see Theorem~3 below).
Figure~\ref{fig:prostate} illustrates the gain in smoothness by calculating SURE for the prostate cancer data
with $P=8$ covariates \citep{Tibs:regr:1996}, and
by comparing the smoothness of the estimated risk either with adaptive lasso (left) 
or with its smooth extension (right).
Both risk estimates are unbiased, but the second is less erratic thanks to $s>1$.
%Visual smoothness would be even more pronounced with data containing a larger collection of covariates.
%The optimal hyperparameters that minimizes SURE is $(\hat \lambda, \hat \nu)=(1.12,10)$ for smooth adaptive lasso,
%which selects a subset with five out of eight covariates.
%Finally we report that the estimated risk with $\nu=1$ (i.e., lasso) is roughly 15, while it drops to roughly 13 with $\nu=10$, showing the potential advantage of smooth adaptive lasso   on this particular data set.

We have assumed unit variance.
In practice, one can either estimate the variance and rescale responses to have approximate unit variance,
or, in the spirit of generalized cross validation \citep{Golu:Heat:Wahb:gene:1979},
one can use generalized SURE
\begin{equation}
{\rm GSURE}(\lambda,\nu;s)=\frac{{\rm RSS}(\hat {\boldsymbol \mu}_{\lambda,\nu;s})/N}{(1-\frac{1}{N}\sum_{n=1}^N \partial g_n({\bf Y};\lambda,\nu,s)/\partial Y_n)^2}.
\label{eq:GSURE}
\end{equation}
% which is a first order approximation to the exact SURE formula and which estimates the variance
% by $\hat \sigma^2={\rm RSS}(\hat {\boldsymbol \mu}_{\lambda,\nu;s})/N$.

To minimize SURE or GSURE over $(\lambda,\nu)$ for $s=2\log \nu +1$, our strategy consists in minimizing it first for $s=1$ (i.e., adaptive lasso)
which can be done efficiently on a fine grid thanks to lars \citep{Efro:Hast:John:Tibs:leas:2004}.
This provides a neighborhood for $\lambda$, namely $[\hat \lambda^{(s=1)}/10, 10 \hat \lambda^{(s=1)}]$,
over which we then minimize SURE on a more local grid
for smooth adaptive lasso ($s>1$) calculated with the SBITE algorithm.
This strategy provides a good and efficient selection of the pair of hyperparameters $(\lambda,\nu)$, as demonstrated by Monte-Carlo below.

% The noise has been assumed to have variance one. In practice, the noise variance must be estimated, for instance based on the residual sum of squares. In that case, the Stein unbiased risk estimate could also take into account this uncertainty, which would entail deriving a more complex SURE formula.

%%%%%%%%%%%%%%%%%%%%%%%%%%%%%%%%%%%%
\subsection{Monte-Carlo simulation}

We replicate the Monte-Carlo simulation of \citet{Zou:adap:2006} with $P=8$ covariates with corresponding coefficients either sparse
$ {\boldsymbol \alpha}=(3,1.5,0,0,2,0,0,0)$ for Model 1 with $N\in \{20,60\}$,
or not sparse $ {\boldsymbol \alpha}=(.85,.85,.85,.85,.85,.85,.85,.85)$ for Model 2  with $N\in \{40,80\}$.
The covariates $\tilde {\bf x}_n$ are i.i.d.~Gaussian vectors with pairwise correlation between $\tilde x_{n,i}$
and $\tilde x_{n,j}$ given by ${\rm cor}(i,j)=(.5)^{|i-j|}$. The noise is Gaussian with standard error $\sigma \in \{1,3,6\}$.
Like \citet{Zou:adap:2006}, we consider the relative prediction error
$
{\rm RPE}={\rm E}[\tilde {\bf x}_{\rm test}\T\{\hat {\boldsymbol \alpha}([\tilde X,{\bf Y}]_{\rm training})-
{\boldsymbol \alpha}\}]/\sigma^2$,
where the expectation is taken over training and test sets, and response.
Note that we exactly calculate RPE given the training set by
using knowledge of the distribution of the covariates (Zou relies on 10,000 test observations instead).
This predictive measure is
reported in Table~\ref{tab:MCsimZou2foldCV} to compare lasso, adaptive lasso and smooth adaptive lasso.
To compare estimators fairly, we consider the same selection rule for all, here two-fold cross-validation.
Since the estimators are used on the same 100 training sets, then the numbers we see in the tables reveal
significant differences, even though marginal standard errors are large.
We observe that SBITE improves
significantly over lasso and adaptive lasso when the underlying model is sparse and the noise is small;
the estimation is slightly worse for the non-sparse model.

\begin{table}[!hb]
\centering
\footnotesize
\caption{Zou's Monte-Carlo simulation with 100 training sets: median RPE using two-fold cross-validation to select
the hyperparameter(s).}\label{tab:MCsimZou2foldCV}
\begin{tabular}{l|rrrrrr} \hline \hline
%              & \multicolumn{3}{c|}{Median RPE} &\multicolumn{3}{c}{Mean RPE} \\
              &  \multicolumn{2}{c}{$\sigma=1$} &  \multicolumn{2}{c}{$\sigma=3$} &  \multicolumn{2}{c}{$\sigma=6$}   \\ \hline
Model 1 & $N=20$ &  $N=60$ & $N=20$ &  $N=60$ &  $N=20$ &  $N=60$ \\
% Lasso$^\dagger$       & $0.414_{(0.046)}$ & $0.395_{(0.039)}$ & $0.275_{(0.026)}$\\
% Adaptive lasso$^\dagger$  & $0.261_{(0.023)}$ & $0.369_{(0.029)}$ & $0.336_{(0.031)}$\\
% SCAD$^\dagger$       & $0.218_{(0.029)}$ & $0.508_{(0.044)}$ & $0.428_{(0.019)}$\\
% Garotte$^\dagger$       & $0.227_{(0.007)}$ & $0.488_{(0.043)}$ & $0.385_{(0.030)}$\\
lasso            & $0.367_{(0.048)}$ & $0.089_{(0.009)}$ & $0.419_{(0.069)}$ & $0.089_{(0.008)}$ &  $0.369_{(0.021)}$ & $0.096_{(0.010)}$\\
adaptive lasso       & $0.360_{(0.051)}$ & $0.052_{(0.009)}$ & $0.435_{(0.057)}$ & $0.085_{(0.009)}$ & $0.308_{(0.021)}$ & $0.097_{(0.011)}$ \\
SBITE & $0.328_{(0.046)}$ & $0.054_{(0.009)}$ & $0.424_{(0.056)}$ & $0.085_{(0.009)}$ & $0.330_{(0.020)}$ & $0.098_{(0.010)}$ \\ \hline
%Model 1 ($N=60$) \\
% Lasso$^\dagger$       & $0.103_{(0.008)}$ & $0.102_{(0.008)}$ & $0.107_{(0.012)}$\\
% Adaptive lasso$^\dagger$       & $0.073_{(0.004)}$ & $0.094_{(0.012)}$ & $0.117_{(0.008)}$\\
% SCAD$^\dagger$       & $0.053_{(0.008)}$ & $0.104_{(0.016)}$ & $0.119_{(0.014)}$\\
% Garotte$^\dagger$       & $0.069_{(0.006)}$ & $0.102_{(0.008)}$ & $0.118_{(0.009)}$\\
%Lasso           & $0.089_{(0.009)}$ & $0.089_{(0.008)}$ & $0.096_{(0.010)}$\\
%Adaptive lasso         & $0.052_{(0.009)}$ & $0.085_{(0.009)}$ & $0.097_{(0.011)}$\\
%Smooth adaptive lasso & $0.054_{(0.009)}$ & $0.085_{(0.009)}$ & $0.098_{(0.010)}$\\
Model 2 & $N=40$ &  $N=80$ & $N=40$ &  $N=80$ &  $N=40$ &  $N=80$  \\
% Lasso$^\dagger$       & $0.205_{(0.015)}$ & $0.214_{(0.014)}$ & $0.161_{(0.009)}$\\
% Adaptive lasso$^\dagger$       & $0.203_{(0.015)}$ & $0.237_{(0.016)}$ & $0.190_{(0.008)}$\\
% SCAD$^\dagger$       & $0.223_{(0.018)}$ & $0.297_{(0.028)}$ & $0.230_{(0.009)}$\\
% Garotte$^\dagger$       & $0.199_{(0.018)}$ & $0.273_{(0.024)}$ & $0.219_{(0.019)}$\\
lasso            & $0.238_{(0.014)}$ & $0.104_{(0.005)}$  & $0.231_{(0.020)}$ & $0.108_{(0.005)}$ &  $0.163_{(0.010)}$ & $0.087_{(0.005)}$\\
adaptive lasso       & $0.238_{(0.015)}$ & $0.104_{(0.005)}$ & $0.233_{(0.021)}$ & $0.108_{(0.005)}$ & $0.181_{(0.010)}$ & $0.091_{(0.005)}$\\
SBITE & $0.238_{(0.015)}$ & $0.104_{(0.005)}$ & $0.240_{(0.021)}$ & $0.109_{(0.005)}$ & $0.172_{(0.010)}$ & $0.090_{(0.005)}$
%Model 2 ($N=80$) \\
% Lasso$^\dagger$       & $0.094_{(0.008)}$ & $0.096_{(0.008)}$ & $0.091_{(0.008)}$\\
% Adaptive lasso$^\dagger$       & $0.093_{(0.007)}$ & $0.094_{(0.007)}$ & $0.104_{(0.009)}$\\
% SCAD$^\dagger$       & $0.096_{(0.104)}$ & $0.099_{(0.012)}$ & $0.138_{(0.014)}$\\
% Garotte$^\dagger$       & $0.095_{(0.006)}$ & $0.111_{(0.007)}$ & $0.119_{(0.006)}$\\
%Lasso       & $0.104_{(0.005)}$ & $0.108_{(0.005)}$ & $0.087_{(0.005)}$\\
%Adaptive lasso       & $0.104_{(0.005)}$ & $0.108_{(0.005)}$ & $0.091_{(0.005)}$\\
%Smooth adaptive lasso & $0.104_{(0.005)}$ & $0.109_{(0.005)}$ & $0.090_{(0.005)}$ \\ \hline
\end{tabular}
%$^\dagger$ results taken from the Monte-Carlo simulation of \citet{Zou:adap:2006}. {\bf In bold}, the best two of six.
%\end{center}
%NOTE: standard error of the order of the precision reported.
\end{table}
\normalsize

We also consider SURE as a selection rule for lasso and its smooth adaptive version SBITE, that we compare
based on their relative prediction error conditional on the covariates of the training set, namely
$
{\rm RPE} \mid \tilde X=\sum_{n=1}^N{\rm E}[\tilde {\bf x}_{{\rm training},n}\T\{\hat {\boldsymbol \alpha}([\tilde X,{\bf Y}]_{\rm training})-
{\boldsymbol \alpha}\}]/\sigma^2$,
where the expectation is taken over the response only. Although less useful in practice unless
prediction is sought at the same locations as the training set, this measure helps quantify the improvement
of using smooth James-Stein thresholding. Table~\ref{tab:MCsimZouSURE} reports the results, which
shows a systematic gain of SBITE over adaptive lasso. Lasso is often better for the non-sparse model.

\begin{table}[!ht]
\centering
%\begin{center}
\footnotesize
\caption{Zou's Monte-Carlo simulation  with 100 training sets: median RPE $\mid X$ at the training covariates $X$ using SURE to select
the hyperparameter(s).}\label{tab:MCsimZouSURE}
\begin{tabular}{l|rrrrrr} \hline \hline
%              & \multicolumn{3}{c|}{Median RPE} &\multicolumn{3}{c}{Mean RPE} \\
      &  \multicolumn{2}{c}{$\sigma=1$} &  \multicolumn{2}{c}{$\sigma=3$} &  \multicolumn{2}{c}{$\sigma=6$}   \\ \hline
Model 1 & $N=20$ &  $N=60$ & $N=20$ &  $N=60$ &  $N=20$ &  $N=60$ \\
lasso                & $0.264_{(0.021)}$ & $0.082_{(0.008)}$ & $0.258_{(0.020)}$ & $0.082_{(0.008)}$ & $0.219_{(0.017)}$ & $0.082_{(0.007)}$  \\         
%Lasso 2-fold CV            & $0.260_{(0.025)}$ & $0.321_{(0.061)}$ & $0.287_{(0.022)}$\\
%Lasso$^\dagger$       & $0.414_{(0.046)}$ & $0.395_{(0.039)}$ & ${\bf 0.275}_{(0.026)}$\\
%Adaptive lasso$^\dagger$       & $0.261_{(0.023)}$ & $0.369_{(0.029)}$ & $0.336_{(0.031)}$\\
%SCAD$^\dagger$       & ${\bf 0.218}_{(0.029)}$ & $0.508_{(0.044)}$ & $0.428_{(0.019)}$\\
%Garotte$^\dagger$       & ${\bf 0.227}_{(0.007)}$ & $0.488_{(0.043)}$ & $0.385_{(0.030)}$\\
adaptive lasso        & $0.231_{(0.023)}$ & $0.075_{(0.008)}$ & $0.280_{(0.023)}$& $0.090_{(0.008)}$ & $0.289_{(0.019)}$ & $0.096_{(0.007)}$\\
SBITE  & $0.228_{(0.023)}$& $0.065_{(0.008)}$ & $0.279_{(0.025)}$ & $0.084_{(0.008)}$ & $0.255_{(0.019)}$ & $0.097_{(0.007)}$ \\ \hline
%Lasso                 & $0.082_{(0.008)}$ & $0.082_{(0.008)}$ & $0.082_{(0.007)}$ \\            
%Lasso 2-fold CV            & $0.078_{(0.008)}$ & $0.077_{(0.007)}$ & $0.083_{(0.010)}$\\
%Lasso$^\dagger$       & $0.103_{(0.008)}$ & $0.102_{(0.008)}$ & $0.107_{(0.012)}$\\
%Adaptive lasso$^\dagger$       & $0.073_{(0.004)}$ & $0.094_{(0.012)}$ & $0.117_{(0.008)}$\\
%SCAD$^\dagger$       & ${\bf 0.053}_{(0.008)}$ & $0.104_{(0.016)}$ & $0.119_{(0.014)}$\\
%Garotte$^\dagger$       & $0.069_{(0.006)}$ & $0.102_{(0.008)}$ & $0.118_{(0.009)}$\\
%Adaptive lasso       & $0.075_{(0.008)}$ & $0.090_{(0.008)}$ & $0.096_{(0.007)}$\\
%Smooth adaptive lasso & $0.065_{(0.008)}$ & $0.084_{(0.008)}$ & $0.097_{(0.007)}$\\
Model 2 & $N=40$ &  $N=80$ & $N=40$ &  $N=80$ &  $N=40$ &  $N=80$  \\
lasso               & $0.187_{(0.010)}$ & $0.092_{(0.004)}$  & $0.187_{(0.010)}$ & $0.090_{(0.004)}$ & $0.137_{(0.008)}$& $0.075_{(0.003)}$ \\    
%Lasso 2-fold CV            & $0.187_{(0.010)}$ & $0.202_{(0.015)}$ & $0.143_{(0.010)}$\\
%Lasso$^\dagger$       & $0.205_{(0.015)}$ & ${\bf 0.214}_{(0.014)}$ & ${\bf 0.161}_{(0.009)}$\\
%Adaptive lasso$^\dagger$       & $0.203_{(0.015)}$ & $0.237_{(0.016)}$ & $0.190_{(0.008)}$\\
%SCAD$^\dagger$       & $0.223_{(0.018)}$ & $0.297_{(0.028)}$ & $0.230_{(0.009)}$\\
%Garotte$^\dagger$       & $0.199_{(0.018)}$ & $0.273_{(0.024)}$ & $0.219_{(0.019)}$\\
adaptive lasso        & $0.191_{(0.011)}$ & $0.092_{(0.004)}$ & $0.237_{(0.013)}$ & $0.127_{(0.006)}$ & $0.172_{(0.009)}$ & $0.105_{(0.004)}$\\
SBITE          & $0.191_{(0.011)}$ & $0.092_{(0.004)}$ & $0.219_{(0.013)}$ & $0.113_{(0.006)}$  & $0.164_{(0.008)}$ & $0.099_{(0.004)}$ 
%Model 2 ($N=80$) \\
%Lasso                & $0.092_{(0.004)}$ & $0.090_{(0.004)}$ & $0.075_{(0.003)}$ \\       
%Lasso 2-fold CV            & $0.092_{(0.004)}$ & $0.096_{(0.004)}$ & $0.081_{(0.004)}$\\
%Lasso$^\dagger$       & $0.094_{(0.008)}$ & ${\bf 0.096}_{(0.008)}$ & ${\bf 0.091}_{(0.008)}$\\
%Adaptive lasso$^\dagger$       & $0.093_{(0.007)}$ & ${\bf 0.094}_{(0.007)}$ & $0.104_{(0.009)}$\\
%SCAD$^\dagger$       & $0.096_{(0.104)}$ & $0.099_{(0.012)}$ & $0.138_{(0.014)}$\\
%Garotte$^\dagger$       & $0.095_{(0.006)}$ & $0.111_{(0.007)}$ & $0.119_{(0.006)}$\\
%Adaptive lasso        & $0.092_{(0.004)}$ & $0.127_{(0.006)}$ & $0.105_{(0.004)}$\\
%Smooth adaptive lasso & $0.092_{(0.004)}$ & $0.113_{(0.006)}$ & $0.099_{(0.004)}$ \\ \hline
\end{tabular}
%$^\dagger$ results taken from the Monte-Carlo simulation of \citet{Zou:adap:2006}. {\bf In bold}, the best two of six.
%\end{center}
%NOTE: standard error of the order of the precision reported.
\end{table}
\normalsize

Finally Table~\ref{tab:MCsimZou2} reports  the number $C$ of selected nonzero components and the number $I$ of zero components incorrectly selected.
We observe that the selection is correct when the noise is small ($\sigma=1$) with SBITE and adaptive lasso
using SURE, but that false detection grows with noise.
\begin{table}[!ht]
\centering
%\begin{center}
\footnotesize
\caption{Median number of Selected Variables for Model 1 with $n=60$}\label{tab:MCsimZou2}
\begin{tabular}{l|ccccc} \hline \hline
& \multicolumn{2}{c}{$\sigma=1$} && \multicolumn{2}{c}{$\sigma=3$} \\
& C & I && C & I \\ \hline
Truth &   3 & 0 && 3 & 0   \\
%Lasso SURE & 5 & 2 & 5 & 2 \\
Lasso$^\dagger$ & 3 & 2 && 3 & 2\\
Adaptive lasso$^\dagger$ & 3 & 1 && 3 & 1\\
SCAD$^\dagger$ & 3 & 0 && 3 & 1\\
Garotte$^\dagger$ & 3 & 1 && 3 & 1.5\\
Adaptive lasso SURE        & 3 & 0 && 4 & 1\\
SBITE SURE & 3 & 0 && 5 & 2 \\
\end{tabular}\\
$^\dagger$ Results taken from the Monte-Carlo simulation of \citet{Zou:adap:2006}.
%\end{center}
%NOTE: standard error of the order of the precision reported.
\end{table}
\normalsize

%%%%%%%%%%%%%%%%%%%%%%%%%%%
%%%%%%%%%%%%%%%%%%%%%%%%%%%

\section{Block canonical regression}
\label{sct:bcp}

We consider block canonical regression, that is when the regression matrix is the identity and the coefficients are organized in blocks of size $Q$, namely,
% \begin{equation}
% Y_{n}^{(q)}=\alpha_{n}^{(q)}+\epsilon_{n}^{(q)}, \quad n=1,\ldots,N, \ q=1\ldots,Q.
% \label{eq:model}
% \end{equation}
\begin{equation}
{\bf Y}_{n}={\boldsymbol \alpha}_{n}+{\boldsymbol \epsilon}_{n} \quad {\rm with} \quad
{\bf Y}_n=\left ( \begin{array}{c} Y_n^{(1)} \\ \vdots \\ Y_n^{(Q)} \end{array} \right ),\ 
{\boldsymbol \alpha}_n=\left ( \begin{array}{c}  \alpha_n^{(1)} \\ \vdots \\ \alpha_n^{(Q)} \end{array} \right )\  {\rm and}\ 
{\boldsymbol \epsilon}_n=\left ( \begin{array}{c} \epsilon_n^{(1)} \\ \vdots \\ \epsilon_n^{(Q)} \end{array} \right ) 
% \quad n=1,\ldots,N, 
\label{eq:model}
\end{equation}
for $n=1,\ldots,N$, where the noise is i.i.d.~Gaussian (independent between  and within sequences).
If the standard deviation is not known, then \citet{Dono94b} proposed an efficient estimate based on the median absolute deviation; another
possibility is to use generalized SURE (\ref{eq:GSURE}).
This setting applies to the gravitational wave bursts detection problem (\ref{eq:Y_j}) of Section~\ref{sct:motiv} with $Q$ captors.
Since $X$ is the identity, rescaling has no impact, that is $\tilde X=X$ and $\beta=\alpha$.

Block sparsity  assumes most blocks ${\boldsymbol \alpha}_n$ are the ${\bf 0}$-vector.
SBITE (\ref{eq:fixedpointdef}) achieves block sparsity and has the closed form expression
% \begin{eqnarray}
% (\hat{\boldsymbol \alpha}_{\lambda,\nu;s})_n &=& \eta_{\lambda,\nu;s}^{SJS} ({\bf Y}_n) \nonumber \\
% &:=& (1-\frac{\lambda^\nu}{\|{\bf Y}_n\|_2^\nu})_+^s {\bf Y}_n \quad \lambda \geq 0,\ \nu >0,\ s \geq 1,\ {\bf Y}_n \in \real^Q,
% \label{eq:SJS}
% \end{eqnarray}
% where $\eta_{\lambda,\nu;s}^{SJS}$ is the smooth James-Stein thresholding function.
\begin{eqnarray}
(\hat{\boldsymbol \alpha}_n)_{\lambda,\nu;s} 
&=& (1-\frac{\lambda^\nu}{\|{\bf Y}_n\|_2^\nu})_+^s {\bf Y}_n, \quad n=1,\ldots,N.
\label{eq:SJS}
\end{eqnarray}
in this canonical setting.
%for $\lambda \geq 0$, $\nu >0$ and $s \geq 1$  for $n=1,\ldots,N$.

\subsection{Total variation of SURE}
\label{subsct:TVSURE}

The Stein unbiased risk estimate also has the closed form expression
%\begin{eqnarray}
${\rm SURE}(\lambda,\nu,s)=\sum_{n=1}^N \hat \rho_n((\lambda,\nu,s),{\boldsymbol \alpha}_n)$ % \label{eq:SURErhon}
%\end{eqnarray}
with $$\hat \rho_n((\lambda,\nu,s),{\boldsymbol \alpha}_n)= \{1-(1-\frac{\lambda^\nu}{\|{\bf Y}_n\|_2^\nu})_+^s\}^2 \|{\bf Y}_n\|_2^2 -Q
+2\sum_{q=1}^Q \frac{\partial (\hat{\boldsymbol \alpha}_n)_{\lambda,\nu;s}}{\partial Y_{n}^{(q)}},$$ where
\begin{equation}
\frac{\partial (\hat{\boldsymbol \alpha}_n)_{\lambda,\nu;s}}{\partial Y_{n}^{(q)}} = \left \{
\begin{array}{ll}
 0 & {\rm if}\ \|{\bf Y}_n\|_2 < \lambda \\
 (1-\frac{\lambda^\nu}{\|{\bf Y}_n\|_2^\nu})^{s-1} (\nu s \lambda^\nu \frac{(Y_{n}^{(q)})^2}{\|{\bf Y}_n\|_2^{\nu+2}} + 1 -\frac{\lambda^\nu}{\|{\bf Y}_n\|_2^\nu} )& {\rm if}\ \|{\bf Y}_n\|_2 \geq \lambda
\end{array}
\right . .
\label{eq:partialalphaSURE}
\end{equation}
For thresholding functions employing no smoothness, that is $s=1$ here, SURE has $N$ discontinuity points as a function of $\lambda$ for a fixed $\nu$. Indeed (\ref{eq:partialalphaSURE}) is discontinuous
at $\lambda=\|{\bf Y}_n\|_2$ for all $n=1,\ldots,N$ when $s=1$, and the size of each jump is equal to $2\nu$.
We had already observed on the left graph of Figure~\ref{fig:prostate} that the larger $\nu$ the more erratic the SURE surface for the prostate cancer data when $s=1$.
There are two negative consequences for the selection of  $\lambda$ and $\nu$.
First the SURE two-dimensional surface will have minima that will be difficult to localize from an optimization point-of-view.
Second, the location of the global minima will be sensitive, in particular with large $\nu$,
to changes in the data ${\bf Y}_n$.

\citet{Dono95i} studied the asymptotic properties of  selecting $\lambda$ by minimizing SURE for $\nu=s=1$.
To show their \emph{SureShrink} estimator is optimally smoothness adaptive,
a key ingredient  is the deviation of SURE around its mean when $\nu=1$.
\begin{figure}[!ht]
  \begin{center}
  \includegraphics[height=9cm, width=14cm]{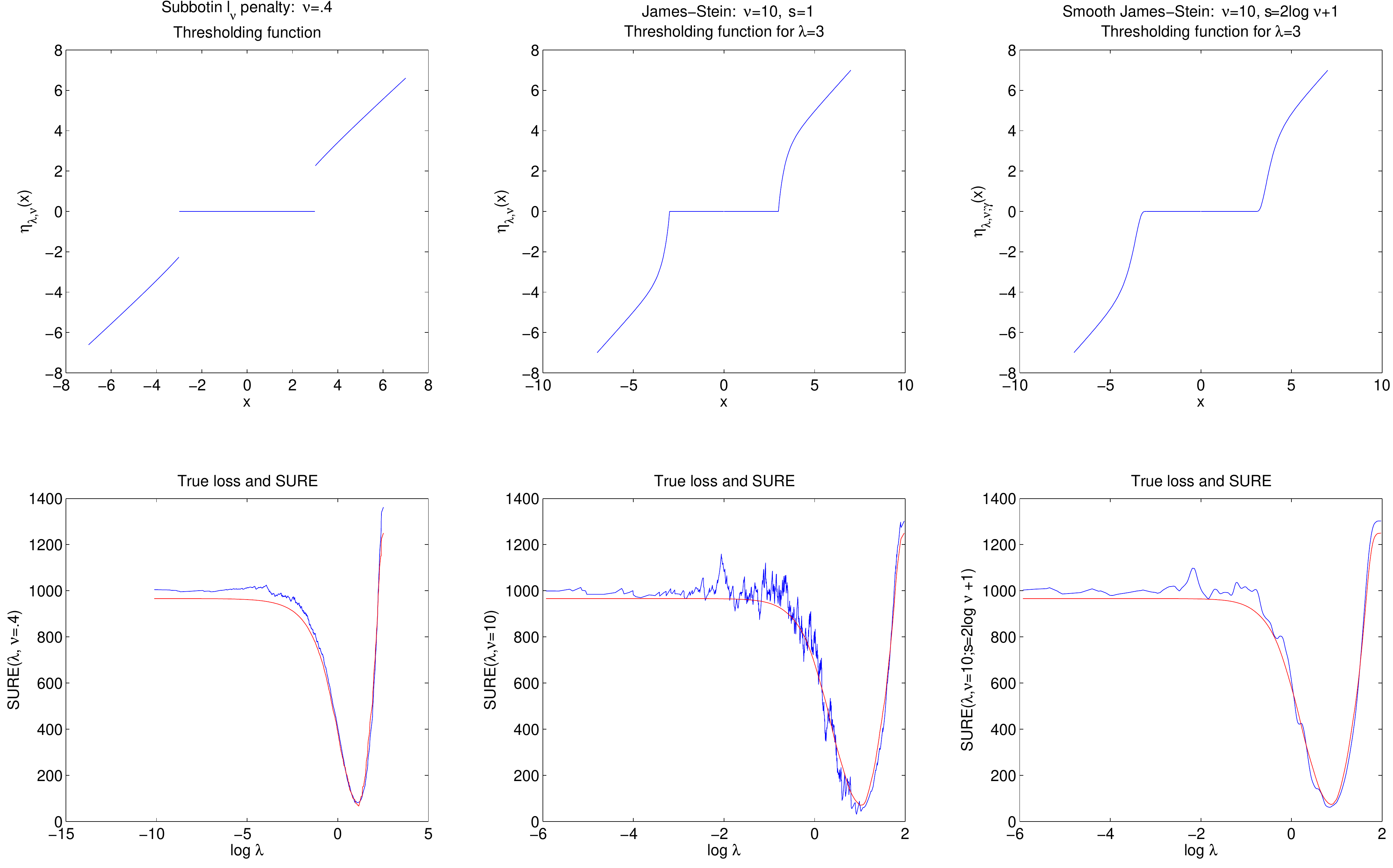}
  \caption{Thresholding functions and corresponding SURE for $Q=1$ and $\nu=10$ on simulated data of length $N=1000$.
Left: Subbotin $\ell_\nu$ penalized least squares; Middle: James-Stein ($s=1$); Right: smooth James-Stein  ($s=2\log \nu+1$).
Top: thresholding functions with parameters chosen to approximate hard thresholding.
The left one is discontinuous at the threshold, but with a small slope at the threshold; the middle one has a discontinuous derivative at the threshold; and the right one has a smooth change of derivative at the threshold.
Bottom: corresponding Stein unbiased risk estimate (least smooth curve) and true loss (smoothest curve).
%Clearly the right estimation ($s>1$) is less erratic than the middle one ($s=1$).}
}
  \label{fig:SURE1-2}
  \end{center}
\end{figure}
Theorem~3 below shows that the total variation of SURE grows  when $\nu$ increases, and that
employing smooth James-Stein thresholding with a smoothness parameter larger than one tempers this erratic effect
by removing the jumps and decreasing the erraticity of SURE.
Figure~\ref{fig:SURE1-2} illustrates the advantage of increasing  $s$ when $\nu$ gets large on simulated data. % of length $N=1000$ for $Q=1$ and $\nu=10$.
% The middle two graphs are for $s=1$: the top graph plots the thresholding function for $\lambda=3$ and the bottom graph shows
% the erratic SURE curve as a function of $\lambda$, along with the smooth true $\ell_2$-loss.
% Likewise the right two graphs are for $s>1$, here $s=2\log \nu+1$.
We observe that while the two thresholding functions for $s=1$ and $s>1$ only differ slightly, the latter is smoother  near the threshold value and
the corresponding SURE curve is less wiggly around the true loss.
Section~\ref{sct:simu} reports results of a Monte-Carlo simulation that quantifies the improvement in mean squared error obtained by adding smoothness.

A measure of erraticity of SURE that is defined not only for $s> 1$ but also for $s=1$ is its total variation as a function of $\lambda$.
The total variation of a function~$f$ in the space of
functions of bounded variation (that is, not necessarily continuous) is 
%\begin{equation}
${\rm TV}(f)=\sup \sum_j |f(\lambda_{j+1})-f(\lambda_j)|$, %, \label{eq:TV1D}
%\end{equation}
where the supremum is taken over all possible partitions
$[\lambda_j,\lambda_{j+1}]$, $j=1,\dots,M$, of the domain of $f$.
(If  $f$ is moreover absolutely continuous,
then TV reduces to the more conventional smoothness measure
${\rm TV}(f)=\int_{\Lambda} |f'(\lambda)|d\lambda$.)
The following theorem quantifies the erraticity of SURE  and shows the tempering effect of the smoothness parameter $s$.
%on the total variation of SURE.

\bigskip
\emph{Theorem~3}: Consider  ${\rm SURE}(\lambda;\nu,s)=\sum_{n=1}^N \hat \rho_n((\lambda;\nu,s),{\boldsymbol \alpha}_n)$ as a function of $\lambda$ for $Q=1$.
%where $\hat \rho_n$ are given in (\ref{eq:SURErhon})
 Its total variation  for a given $\nu \geq 1$  satisfies
$$
{\rm TV}^{(s=1)}({\rm SURE})=\sum_{n=1}^N {\rm TV}^{(s=1)}(\hat \rho_n)  > \sum_{n=1}^N {\rm TV}^{(s>1)}(\hat \rho_n) \geq {\rm TV}^{(s>1)}({\rm SURE}).
$$
Moreover erraticity increases less with $s>1$ when  $\nu$ or $|Y_n|$ grows since
$\frac{\partial}{\partial \nu}  {\rm TV}^{(s>1)}(\hat \rho_n)  \leq \frac{\partial}{\partial \nu}  {\rm TV}^{(s=1)}(\hat \rho_n)$
and  $\frac{\partial}{\partial Y_n}  {\rm TV}^{(s>1)}(\hat \rho_n) \leq \frac{\partial}{\partial Y_n}  {\rm TV}^{(s=1)}(\hat \rho_n)$ for ${Y_n \geq 0}$.
In particular when $Y_n \rightarrow 0$ and for $\nu$ large, then $\frac{\partial}{\partial \nu}  {\rm TV}^{(s>1)}(\hat \rho_n) \rightarrow  4 (1-1/s)^{s-1} \geq 4 \exp(-1)$
for $s$ fixed. Letting $s$ grow slowly with $\nu$, for instance $s(\nu)=2 \log \nu+1$, then the lower bound $4 \exp(-1)$ is reached to lower erraticity most.

\subsection{Universal threshold and information criterion}
\label{subsct:SLIC}

To derive a universal threshold \citep{Dono94b} and an information criterion  for SBITE,
we approximate below the distribution of the smallest threshold $\lambda_{{\cal Y}}$ that, for a sample ${\cal Y}=({\bf Y}_1, \ldots, {\bf Y}_N)$ of size $N$,
sets to zero all  $N$ blocks of length $Q$  when the true underlying model is made of zero vectors.
Controlling the maximum of $\lambda_{{\cal Y}}$ then leads to a finite sample $\tilde \lambda_{N,Q}$ and asymptotic $\lambda_{N,Q}$ universal thresholds, 
and a prior distribution $\pi_\lambda$ for $\lambda$.
%used to define the information criterion. % as for the sparsity $\ell_\nu$ information criterion SL$_\nu$IC \citep{SardySLIC09}.

Assuming ${\bf Y}_n\stackrel{{\rm i.i.d.}} \sim {\rm N}_Q({\bf 0}, I_Q)$ for $n=1,\ldots,N$, we seek the smallest threshold $\lambda_{N,Q}$
such that SBITE estimates the right model with a probability tending to one:
\begin{equation}
{\rm P}((\hat{\boldsymbol \alpha}_1)_{\lambda_{N,Q},\nu;s}={\bf 0}, \ldots, (\hat{\boldsymbol \alpha}_N)_{\lambda_{N,Q},\nu;s}={\bf 0}) =
{\rm P}(\max_{n=1,\ldots,N }\|{\bf Y}_n\|_2^2 \leq \lambda_{N,Q}^2) 
\stackrel{N \rightarrow \infty}\longrightarrow  1.
\label{eq:univ}
\end{equation}
The distribution of $M_N=\max_{n=1}^N \|{\bf Y}_n\|_2^2$, where $\|{\bf Y}_n\|_2^2\stackrel{\rm i.i.d.} \sim \chi_Q^2=\Gamma(Q/2,1/2)$, is degenerate. Extreme value theory provides proper rescaling of $M_N$ for
$c_N^{-1} (M_N-d_N(Q))\longrightarrow_d G_0(x)$,
where $G_0(x)=\exp(-\exp(-x))$ is the Gumbel distribution, $c_N=2$ and $d_N(Q)$ is the root in $\xi$ to
\begin{equation}
\log N-\log \Gamma(Q/2)=(1-Q/2)\log (\xi/2)+\xi/2.
\label{eq:rootdN}
\end{equation}
The normalizing constant $d_N(Q)=2(\log N+(Q/2-1) \log \log N - \log \Gamma(Q/2))$  given by \citet[p.156]{Embr:Klup:Miko:mode:1997} for the Gamma distribution
is the asymptotic root of (\ref{eq:rootdN}), which provides a good Gumbel approximation when $N$ is large compared to $\Gamma(Q/2)$.
In that case we define the asymptotic universal threshold
%\begin{equation}
$\lambda_{N,Q}=\sqrt{2(\log N+(Q/2) \log \log N - \log \Gamma(Q/2))}$,
%\label{eq:asymlambdaNQ}
%\end{equation}
for which~(\ref{eq:univ}) is satisfied since
\begin{equation}
{\rm P}(\max_{n=1,\ldots,N }\|{\bf Y}_n\|_2^2 \leq \lambda_{N,Q}^2)\stackrel{\cdot}= G_0(\log \log N)\approx 1-1/\log N \stackrel{N \rightarrow \infty} \longrightarrow 1.
\label{eq:converge}
\end{equation}
Note that  we get back the standard universal threshold $\sqrt{2\log N}$ up to a small term for $Q=1$, and
the universal threshold of \citet{minimaxC:2000} for denoising complex-valued signals for $Q=2$.
When $Q$ gets large however, the proposed normalizing constant $d_N(Q)$  is too far from the exact root to provide a useful approximation,
so we  find the root $d_N(Q)$ of (\ref{eq:rootdN}) numerically. %, which exists provided $N$ is large enough (e.g., $N>\log(\Gamma(Q/2))+(Q/2-1)(1-\log(Q/2-1))$).
The finite sample universal threshold is then defined as
\begin{equation}
\tilde \lambda_{N,Q}=\sqrt{d_N(Q)+c_N \log \log N} \quad \mbox{ with } \quad d_N(Q)\mbox{ root of } (\ref{eq:rootdN})
\label{eq:lambdaNQ}
\end{equation}
 to have the same rate of convergence  for all $Q$ with $\tilde \lambda_{N,Q}$ in place $\lambda_{N,Q}$ in (\ref{eq:converge}).

% {\bf Proof}: Studying (\ref{eq:rootdN}) as a function of $\xi$, one sees that the root for a finite sample is larger than the asymptotic root given by \citet[p.156]{Embr:Klup:Miko:mode:1997} provided $(Q/2-1) \log \log N-\log \Gamma(Q/2)\geq 0$. %The the second inequality also holds from (\ref{eq:asymlambdaNQ}).

More than a bound the asymptotic Gumbel pivot for $M_N$ leads to a prior distribution 
$F_{\lambda}(\lambda)=G_0((\lambda^2-d_N(Q))/2)$ of the threshold $\lambda$ to reconstruct true zero vectors from noisy measurements.
When $s=1$, Bayes theorem provides the joint posterior distribution of the coefficients and the hyperparameters. Taking its negative logarithm leads to the following information criterion in the spirit of the sparsity $\ell_\nu$ information criterion SL$_\nu$IC \citep{SardySLIC09}.

\bigskip
\emph{Definition.} \label{def:SWLIC}  Suppose model (\ref{eq:model}) or model (3.1) of \citet{Cai:adap:1999} holds.
The sparsity weighted $\ell_2$ information criterion for the estimation of $({\boldsymbol \alpha}_1, \ldots, {\boldsymbol \alpha}_N)$ and
the selection of $(\lambda,\nu)$ with SBITE (\ref{eq:SJS}) for $s=1$ is defined as
\begin{eqnarray*}
{\rm SL}_2^w{\rm IC}({\boldsymbol \alpha}_1, \ldots, {\boldsymbol \alpha}_N, \lambda, \nu)&=&  \frac{1}{2} \sum_{n=1}^N \|  {\bf Y}_n- {\boldsymbol \alpha}_n \|_2^2 + \lambda^\nu \sum_{n=1}^N \frac{1}{\|{\bf Y}_n \|_2^{\nu-1}} \| {\boldsymbol \alpha}_n \|_2   \nonumber \\
&&  - N \log(\frac{\Gamma(Q/2)}{2\pi^{Q/2}\Gamma(Q)}) + Q(\nu-1)\sum_{n=1}^N \log \|{\bf Y}_n\|_2 \nonumber \\
&&-QN\nu\log \lambda   -\log \pi_{\lambda}(\lambda;\tau_{N,Q})-\log \pi_\nu(\nu), \nonumber
%\label{eq:SWLIC}
\end{eqnarray*}
where $\pi_\nu$ is a prior for $\nu$ that we choose Uniform on $[1,\infty)$, $\pi(\lambda; \tau)=F'(\lambda; \tau)$ with $F_{\lambda}(\lambda;\tau)=G_0((\lambda^2/\tau^2-d_N(Q))/2)$ and $\tau$ is calibrated to
$\tau_{N,Q}^2=\tilde \lambda_{N,Q}^2/(QN\nu+1)$ to match the asymptotic model consistency %when the true sequences are null, i.e.,
when ${\boldsymbol \alpha}_n={\bf 0}$ for $n=1,\ldots,N$.

\bigskip
In practice,  one minimizes SL$_2^w$IC like AIC or BIC to both select the hyperparameters $(\lambda,\nu)$ and estimate the sequences ${\boldsymbol \alpha}_n$, $n=1,\ldots,N$.
The information criterion could also be derived for $s>1$ if we knew the definition of SBITE as a penalized least squares, which is an open problem.

\subsection{Oracle inequality}
\label{sct:oracle}

% Interestingly the James-Stein estimator has the following oracle inequality
% \begin{equation}
% {\rm R}(\hat {\boldsymbol \alpha}^{{\rm JS+}},{\boldsymbol \alpha})\leq 2 + \inf_{c\in \real} {\rm R}(\hat {\boldsymbol \alpha}^{c},{\boldsymbol \alpha}),
% \label{eq:oracle1}
% \end{equation}
% with respect to the linear estimator $\hat {\boldsymbol \alpha}^{c}=c\hat {\boldsymbol \alpha}^{{\rm MLE}}$, for which the oracle factor is $c^*=\|{\boldsymbol \alpha} \|_2^2/(P+\|{\boldsymbol \alpha} \|_2^2)$ and the optimal $\ell_2$-risk is ${\rm R}^*(c,{\boldsymbol \alpha})=\inf_c {\rm R}(\hat {\boldsymbol \alpha}^{c},{\boldsymbol \alpha})=P\|{\boldsymbol \alpha} \|_2^2/(P+\|{\boldsymbol \alpha} \|_2^2)$.
\cite{Candes:2005} provides an interesting review on oracle inequalities. Here we derive an oracle inequality for SBITE employing smooth James-Stein
thresholding when the block size $Q\geq 2$ is fixed.
\citet{Cai:adap:1999} derived an oracle inequality for block sizes increasing with the sample size.
The $\ell_2$ risk for model (\ref{eq:model}) is 
 %\begin{eqnarray*}
$
{\rm R}(\hat{\boldsymbol \alpha},{\boldsymbol \alpha})=\sum_{n=1}^N \rho_n({\boldsymbol \alpha}_n)=\sum_{n=1}^N {\rm E} \|\hat {\boldsymbol \alpha}_{n} -  {\boldsymbol \alpha}_n\|_2^2
$. %\end{eqnarray*}
Following \citet{Dono94b}, 
the oracle predictive performance of the block diagonal projection estimator $\hat {\boldsymbol \alpha}_n=\delta_n {\bf Y}_n$, where $\delta_n \in \{0,1\}$ is
$$
\rho_n(\delta_n,{\boldsymbol \alpha}_n)=\left \{  \begin{array}{ll}
\| {\boldsymbol \alpha}_n \|_2^2, & {\rm if} \ \delta_n=0, \\
Q,  & {\rm if} \ \delta_n=1.
\end{array} \right . 
$$
Hence the oracle hyperparameters are $\delta_n^*=1_{\{\| {\boldsymbol \alpha}_n\|_2^2>Q\}}$ for $n=1,\ldots,N$,
and the corresponding oracle overall risk  is
$
{\rm R}^*({\boldsymbol \delta},{\boldsymbol \alpha})=\sum_{n=1}^N \min(\|{\boldsymbol \alpha}_n \|^2,Q)
$.
The following theorem extends  the oracle inequality obtained by \citet{Dono94b} and \citet{Zou:adap:2006} for $Q=s=1$ to block thresholding with $Q\geq 2$ and $s>1$.

\bigskip
\emph{Theorem~4}: For any fixed $Q\geq 2$, there exists a sample size $N_0$ such that, for all $N\geq N_0$ and with the universal threshold $\tilde \lambda_{N,Q}$
defined in (\ref{eq:lambdaNQ}), then SBITE defined by (\ref{eq:SJS}) for $\nu\geq 1$ and $s\geq 1$ achieves the oracle inequality
\begin{equation}
{\rm R}(\hat{\alpha}^{{\rm SBITE}}_{\tilde \lambda_{N,Q},\nu;s},{\boldsymbol \alpha})\leq (Q+1+2\nu s+ c_{\nu,s,Q} \lambda_{N,Q}^2)(Q+{\rm R}^*({\boldsymbol \delta},{\boldsymbol \alpha})),
\label{eq:OrIneq}
\end{equation}
where $c_{\nu,s,Q}=\max(1+\frac{\nu s}{Q},s^2)$ and $ \lambda_{N,Q}^2= 2 \log N+Q\log \log N-2\log \Gamma(Q/2)$.

\bigskip

This result shows we can mimic the overall oracle risk achieved with $N$ oracle hyperparameters within a factor of essentially
$\lambda_{N,Q}^2$ with the single hyperparameter $\tilde \lambda_{N,Q}$.
The smallest sample size $N_0$ for which the inequality holds is quite small in practice; more work is needed to get a tight expression.
Note that  for $s=1$, the inequality differs from \citet{Zou:adap:2006} which had the $\nu$-term in the denominator (this seems to be due to an error in
 $d\hat\mu_i^*(\lambda)/d y_i$ right above (A.13) p.~1427).
This result shows that increasing $\nu$ or $s$ increases the oracle inequality constant. But this does not prevent
the estimator with  $\nu> 1$ to be oracle \citep{Zou:adap:2006}, which is  not  true for lasso with $\nu=1$.
Likewise using a larger $s$  improves
predictive performance in practice although the oracle inequality constant increases.
%compared to the estimator with the parameters set to their lower bound, namely $\nu=s=1$.

%%%%%%%%%%%%%%%%%%%%%%%%%%%
%%%%%%%%%%%%%%%%%%%%%%%%%%%
\subsection{Application to wave burst detection and estimation}
\label{sct:appli}

We employ  SBITE  blockwise and levelwise to $Q=3$ concomitant time series of length $T=2^{14}$ (about 3.27 seconds of recording)
to detect gravitational wave bursts, as described in Section~\ref{sct:motiv}.
\begin{figure}[!h]
  \begin{center}
% \centerline{\includegraphics[height=6in,width=3in]{paperfig1}}
\includegraphics[height=30cm, angle=-90,width=15.0cm]{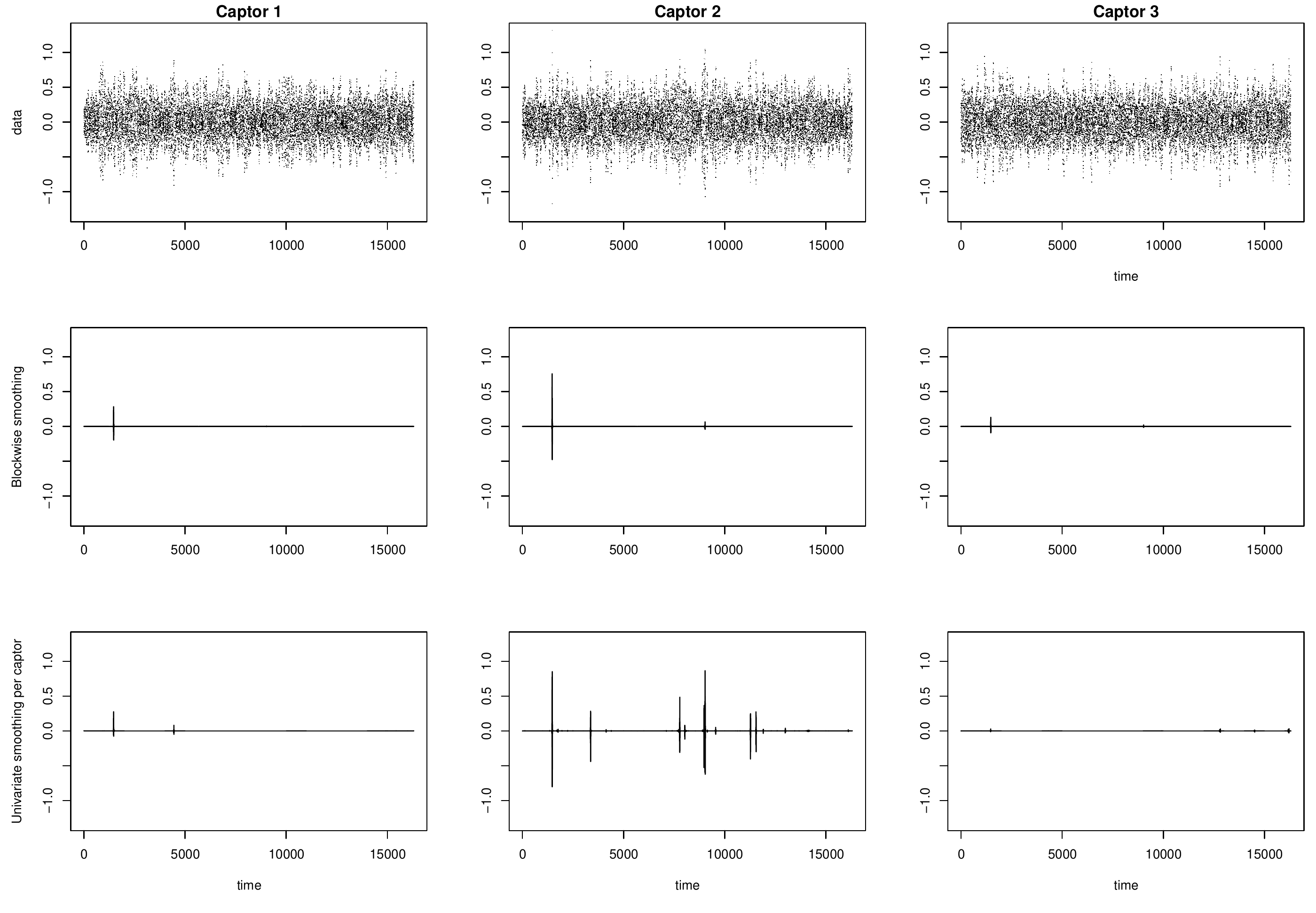}
  \caption{A wave burst injection at time $t=1500$ added on three real captors noise (columnwise).
Top: Time series of length $T=2^{14}$.
Middle: SBITE employing SURE levelwise and blockwise.
Bottom: univariate smoothing per captor levelwise.}
  \label{fig:global}
  \end{center}
\end{figure}
Taking $J=4$ in the wavelet expansion (\ref{eq:wavelet-based-expansion}),
SBITE has  a total of $22$ hyperparameters  to select (11 levels with two hyperparameters each).
Data are pure electronic noise, so we add three proportional so-called ``injections'' at time $t=1500$,
to mimic a wave burst.

\begin{figure}[!hb]
  \begin{center}
% \centerline{\includegraphics[height=6in,width=3in]{paperfig1}}
\includegraphics[height=30cm, angle=-90,width=15.0cm]{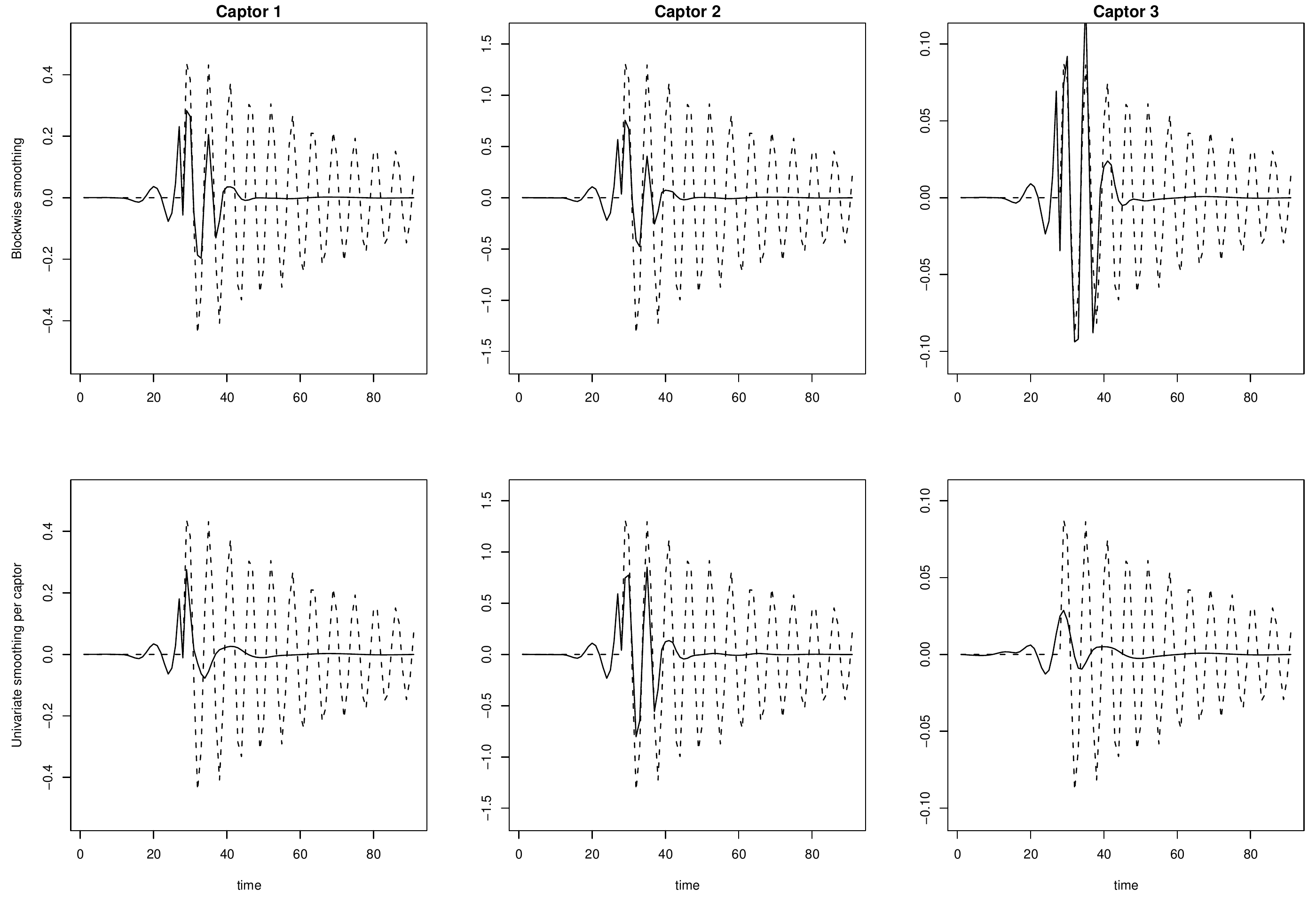}
  \caption{Zooming on the ``injection'' represented by the dotted line (the $y$-scales differs between captors).
The estimates are plotted with a continuous line: blockwise smoothing (first line),
and  a univariate smoothing per captor (second line).}
  \label{fig:local}
  \end{center}
\end{figure}

Figure~\ref{fig:global} shows the data (first line) for each captor (columnwise), the SBIT estimate (second line)
and estimates employing a univariate smoothing per captor (third line).
We observe that, as opposed to univariate smoothing, blockwise smoothing detects the injection and has no false detection except right before time $t=10'000$.
Figure~\ref{fig:local} zooms around the injections that are three times larger on the second captor, and five times smaller on the third captor.
We see that  blockwise estimation of the injections is better than coordinatewise. 

%%%%%%%%%%%%%%%
\subsection{Monte-Carlo simulation}
\label{sct:simu}

We reproduce the Monte-Carlo simulation of \citet{JS04} for $Q=1$ captor to estimate a sparse sequence of length $N=1000$ and of varying degrees of sparsity, as measured by the number of nonzero terms taken in $\{5,50,500\}$ and by the value of the nonzero terms $\mu$ taken in $\{3,4,5,7\}$.
Table~\ref{tab:MCsim1} reports  estimated risks of four estimators: SBITE with ($s>1$) and without ($s=1$) smoothness,
Subbotin $\ell_\nu$ penalized likelihood \citep{SardySLIC09} and EBayesThresh \citep{JS04}. 
The results clearly show the superiority of SBITE with SURE thanks to more smoothness.
%Indeed smoothness of the estimator and of the SURE surface with $s>1$ always improves SBITE. 
In case of extreme sparsity, only SBITE  using the SL$_2^w$IC information criterion and EBayesThresh perform better; 
this drawback of SURE has been explained by \citet[Section 2.4]{Dono95i}.

\begin{table}[!ht]
\centering
\scriptsize
%\begin{center}
\caption{Monte-Carlo simulation for a sequence of length $N=1000$.
Average total squared loss of: SBITE using smooth SURE; SBITE with smoothness parameter fixed to $s=1$ using SURE or SL$_2^w$IC; 
the Subbotin$(\lambda,\nu)$ posterior mode estimator using SURE or SL$_\nu$IC; 
and the EBayesThresh estimator with Cauchy-like prior. %In {\bf bold}, the best between all methods for each loss.
}\label{tab:MCsim1}
\begin{tabular}{lrrrrrrrrrrrrrr} \hline \hline
Number nonzero & \multicolumn{4}{c}{5} &&\multicolumn{4}{c}{50} &&
\multicolumn{4}{c}{500}  \\ \cline{2-5} \cline{7-10} \cline{12-15}
 Value nonzero $\mu=$ & 3 & 4& 5 &7          && 3 & 4& 5 &7 && 3 & 4& 5 &7 \\ \hline
\underline{$Q=1$ captor}\\
% Laplace $(w,a)$\\
%\ \ $\ell_2$ loss & {\bf 35} & {\bf 33} & {\bf 19} & {\bf 9}   && {\bf 211} & {\bf 154} & {\bf 102} & 72 && 856 & 873 & 782 & 661 \\
%\ \ $\ell_1$ loss & {\bf 13} & 11 &  8 & {\bf 5}   &&  95 &  74 &  59 & 49 && 709 & 721 & 620 & 502\\ 
%\ \  Type I       & 2  & 1  &  1 & 0.5 &&  16 &  12 &   8 &  4 && 500 & 500 & 310 & 98 \\
%\ \  Type II      & 3  & 1  &0.3 & 0   &&  14 &   3 & 0.6 &  0 &&   0 & 0   & 0 &  0\\
SBITE $s>1$\\
\ \ SURE    & 37 & 37 & 26 & 15   && 202 & 165 & 109 & 65 && 826 & 752 & 614 & 521 \\
%\ \ $\ell_1$ loss & 28 & 26 & 18 & 10   &&  137 & 94 &  67& 48 &&  676 & 510 & 442 & 407 \\
SBITE $s=1$ \\
\ \ SURE    & 45 & 42 & 31 & 24   && 213 & 174 & 119 & 77 && 848 & 760  & 624 & 551 \\
%\ \ $\ell_1$ loss & 26 & 22 & 16 & 13   &&  124 &  93 &  68& 52 &&  660 & 507 & 445 & 419 \\
\ \ SL$_2^w$IC   & 39  & 41  & 23  & 6     && 380 & 389 & 213 & 54 && 3350 & 2688 & 1532 & 532 \\
%\ \ $\ell_1$ loss & 13  & 12  & 7  &  4     && 132 & 115  &  69 & 40 && 1204 & 876 &  578 & 403 \\
Subbotin\\
\ \ SURE    & 39 & 35 & 27 & 25 && 232 & 167 & 107 & 97 && 1239& 794 & 607 & 533 \\
%\ \ $\ell_1$ loss & 17 & 14 & 12 & 12           &&  95 &  71 &  57 & 56 && 589 & 468 & 431 & 410 \\
\ \ SL$_\nu$IC    & 38  & 36  & 19  & 9         && 356 & 296 & 132 & 59 && 849 & 831 & 839 & 859 \\
%\ \ $\ell_1$ loss & 13  & 10  &  7  & 5         && 126 & 89  &  54 & 43 && 636 & 654 & 671 & 696 \\ 
EBayesThresh & 37  & 36  & 19  &  8   && 268 & 177 & 104 & 77 && 924 & 899 & 831 & 743 \\
%\ \  $\ell_1$ loss & 13  & 11 &  7 & 5      && 102 &  74 &  58 & 53 && 704 & 683 & 658 & 628\\ \hline
\underline{$Q=3$ captors}\\
SBITE  $s>1$ & 18 & 17 & 14 & 8.7   && 106 & 94 & 75 & 56 && 615 & 615 &  558 & 510 \\
%\ \ $\ell_1$ loss & 17 & 14 & 11 & 6.8   &&  84 & 66 & 53 & 44 && 537 & 482 &  427 & 404 \\
SBITE $s=1$ \\
\ \ SURE  & 22 & 21 & 18 & 16   && 110 & 98 & 80 & 63  && 612 & 613 &  563 & 516 \\
% \ \ $\ell_1$ loss & 16 & 13 & 11 & 10   &&  78 &  65 &  54 & 47 && 518 & 469 &  429 & 407 \\
\ \ SL$_2^w$IC   & 19  & 18  & 11  & 5.7        && 181 & 168 & 109 & 52 && 1600 & 1350 & 807 & 548 \\
%\ \ $\ell_1$ loss & 8.6  & 7.4  & 5.5  &  4.2     && 83 & 71  &  53 & 40 && 770 & 624 & 573 & 414 \\ \hline
\end{tabular}
%\end{center}
%NOTE: standard error of the order of the precision reported.
\end{table}
\normalsize

We also perform a Monte-Carlo simulation with now $Q=3$ concomitantly observed sequences: all three underlying sequences are identical in the location 
of the non zero entries, but not in their amplitude. Since three sequences carry more information than a single one, we may hope to distinguish noise from signal with a smaller
signal-to-noise ratio, so we consider
value of the nonzero terms fixed to $\mu_1=1$, $\mu_2=2$ and $\mu_3=\mu$ taken in   $\{3,4,5,7\}$. 
% We compare the smooth James-Stein estimator (with SURE, $s\geq 1$) and the generalized James-Stein estimator (with SL$_2^w$IC and SURE, $s=1$)  for a single estimation taking all three sequences at once.
 The estimated risks (divided by $Q$ to allow some comparison with $Q=1$) 
are reported in Table~\ref{tab:MCsim1}.
SBITE with smoothness ($s>1$) again performs best overall, while the SL$_2^w$IC selection rule  performs better than for $Q=1$,
 even more so with very sparse sequences.

%%%%%%%%%%%%%%%%%%%%%%%%%%%%
\section{Further extensions}
\label{sct:further}

Smooth James-Stein thresholding (\ref{eq:SJS}) relies on the $\ell_2$ norm, like most block thresholding we are aware of.
This measure may not be appropriate in certain applications.
Indeed if one wants to measure departure from the zero vector in a sense that all entries must be different from zero,
then $\|{\bf Y}_n \|_2$ in (\ref{eq:SJS}) should be replaced by  $\min_{q=1,\ldots,Q} |Y_{n}^{(q)}|$ leading to 
robust SBITE:
$$
(\hat{\boldsymbol \alpha}_n)_{\lambda,\nu;s} = (1-\frac{\lambda^\nu}{\min_{q=1,\ldots,Q} |Y_{n}^{(q)}|^\nu})_+^s {\bf Y}_n;
$$
%\begin{eqnarray}
%\hat{\boldsymbol \alpha}_{n,\lambda,\nu;s} &=& (1-\frac{\lambda^\nu}{\min_{q=1,\ldots,Q} |Y_{nq}|^\nu})_+^s {\bf Y}_n \quad \lambda \geq 0,\ \nu >0,\ s \geq 1,\ {\bf Y}_n \in \real^Q,
%\label{eq:minSJS}
%\end{eqnarray}
a simple calculation leads to $\lambda_{N,Q}=\sqrt{2/Q \log N}$ for its  corresponding  universal threshold.
Other quantiles, or a norm like the $\ell_1$ norm, could also be considered.
Importantly also, the use of a common threshold $\lambda$ for each block implicitly assumes blocks of equal size; if not, then the threshold $\lambda(p_j)$
must grow with block size $p_j$.

For the group lasso, \citet{Yuan:Lin:mode:2006} provided an approximate degrees of freedom.
One can instead follow the derivation of Theorem~2 to derive the exact one, as we did in Section~\ref{subsct:edf} for adaptive lasso. Consider
the smooth  generalization of adaptive group lasso (\ref{eq:adaptlasso}), defined as solution to
% \begin{eqnarray}
%  \hat {\boldsymbol \beta}_j({\bf Y}) &=& {\bf s}_j
% \left \{\frac{\eta^{\rm soft}_{\lambda^\nu}(\|\hat {\boldsymbol \beta}_j^{\rm LS}\|_2^{\nu-1}\|{\bf s}_j \|_2)}
% {\|\hat {\boldsymbol \beta}_j^{\rm LS}\|_2^{\nu-1}   \|{\bf s}_j\|_2}\right \}^s \nonumber \\
% && {\rm with} \quad {\bf r}_{-j}={\bf Y}-X_{-j}\hat {\boldsymbol \beta}({\bf Y}) \quad {\rm and} \quad {\bf s}_j={\bf X}_j\T {\bf r}_{-j} \quad j=1,\ldots,J,
% %p=1,\ldots,P, % \quad {\rm with} \quad {\bf r}_{-p}={\bf Y}-X_{-p} \hat {\boldsymbol \alpha}({\bf Y}).
% %\hat \alpha_p({\bf Y}) &=& \frac{1}{\|{\bf x}_p \|_2^2} \eta^{\rm SJS}_{\lambda,\nu;s}(({\bf Y}-X_{-p}\hat {\boldsymbol \alpha}({\bf Y}))^{\rm T} {\bf x}_p), \quad p=1,\ldots,P. \label{eq:impliciteSJS}
% % &=& g_p({\bf Y})+Y_p  \quad p=1,\ldots,P
% \label{eq:group}
% \end{eqnarray}
\begin{eqnarray}
 \hat {\boldsymbol \beta}_j({\bf Y}) &=&  (c_j)_+^s {\bf s}_j \quad 
{\rm with} \quad \left  \{ \begin{array}{l}
c_j=1-\lambda^\nu /( \|\tilde {\boldsymbol \beta}_j^*\|_2^{\nu-1}   \|{\bf s}_j\|_2)   \\
%{\bf s}_j={\bf X}_j\T \{ {\bf Y}-X_{-j}\hat {\boldsymbol \beta}({\bf Y})\} \\
{\bf s}_j= X_j\T {\bf Y}-\sum_{k\neq j} X_j^T X_k \hat {\boldsymbol \beta}_k 
 \end{array} \right . j=1,\ldots,J,
%p=1,\ldots,P, % \quad {\rm with} \quad {\bf r}_{-p}={\bf Y}-X_{-p} \hat {\boldsymbol \alpha}({\bf Y}).
%\hat \alpha_p({\bf Y}) &=& \frac{1}{\|{\bf x}_p \|_2^2} \eta^{\rm SJS}_{\lambda,\nu;s}(({\bf Y}-X_{-p}\hat {\boldsymbol \alpha}({\bf Y}))^{\rm T} {\bf x}_p), \quad p=1,\ldots,P. \label{eq:impliciteSJS}
% &=& g_p({\bf Y})+Y_p  \quad p=1,\ldots,P
\label{eq:group}
\end{eqnarray}
where the coefficients vector  is segmented as ${\boldsymbol  \beta}=({\boldsymbol  \beta}, \ldots, {\boldsymbol  \beta}_J) \in \real^P$
in $J$ blocks of respective size $p_j$ such that $\sum_{j=1}^J p_j=P$.
%Likewise let $X=[X_1 \ldots X_J]$ with $X_j\T X_j = I_{p_j}$ for $j=1,\ldots,J$.
Note that (\ref{eq:group}) is the smooth extension  of  \citet[(2.4)]{Yuan:Lin:mode:2006} taking $K_j=I_{p_j}$ using their notation.
%Let ${\cal J}_0=\{j\in\{1,\ldots,J\}: \hat {\boldsymbol \beta}_j({\bf Y})= {\bf 0}\}\}$.
Deriving SURE in that setting requires calculating the nonzero elements %${\bf h}_n^{\bar {\cal J}_0}$
 of the block gradient
$\nabla_n \hat {\boldsymbol \beta}({\bf Y})$ of the estimated vector  defined by (\ref{eq:group}) with respect to the data $Y_n$ for $n=1,\ldots,N$.
Following similar derivation as in Appendix~\ref{app:th2}, one finds they %${\bf h}_n^{\bar {\cal J}_0}$
are defined by a set of linear equations of the form (\ref{eq:lineqgrad}), that has a unique solution when $s>1$.
Hence the exact equivalent degrees of freedom of smooth adaptive group lasso can be calculated, and in particular for adaptive lasso
by letting the smoothness parameter tend to one.

%, more precisely
%\begin{equation}
% C_j {\bf X}_{j}\T X^{\bar {\cal J}_0} D_{j}^{\bar {\cal J}_0} {\bf h}_n^{\bar {\cal J}_0}={\bf z}_{n,j}\quad \mbox{for all}\quad  j=1,\ldots,\bar {\cal J}_0,
%\label{eq:SUREgroup}
%\end{equation}
%where ${\bf z}_{n,j}$ are vectors of length $p_j$ and $D_{j}^{\bar {\cal J}_0}$ is a block diagonal matrix made of identity matrices,
%except for the $j$th block equal to $C_j^{-1}$, with
%$C_j=e_j {\bf s}_j{\bf s}_j\T+f_j I_{p_j}$ is a positive definite matrix since
%$e_j=s (1/w_j)^{s-1} \lambda^\nu/\|\tilde {\boldsymbol \beta}_j^*\|_2^{\nu-1}/\|{\bf s}_j\|_2^3>0$
%and $f_j=(1/w_j)^{s}>0$.
%Based on (\ref{eq:SUREgroup}), SURE can be calculated and minimized over all values of $s > 1$ and $\nu \geq 1$ to provide the equivalent
%degrees of freedom of  (smooth adaptive) group lasso.

%Unless blocks are of unit size,  there does not seem to be a close form expression
%for the degrees of freedom of group lasso when $\nu=1$ and $s \rightarrow 1$. 

%$$
%\min_{{\boldsymbol \alpha}_1, \ldots, {\boldsymbol \alpha}_J} \frac{1}{2} \|{\bf y}- X {\boldsymbol \alpha} \|_2^2
%$$

%%%%%%%%%%%%%%%%%%%%
%%%%%%%%%%%%%%%%%%%%
\section{Conclusions}

We developed SBITE variable selection defined as the fixed point of an iterative sequence employing the smooth James-Stein thresholding function.
SBITE can be employed blockwise or coordinatewise, and can control sparsity, shrinkage and smoothness by means of three parameters.
For any combination of these three parameters, we have derived the Stein unbiased risk estimate 
that is smoother the larger $s$ for a better selection of the regularization parameters.
Letting the smoothness parameter tend to one, we obtained the equivalent degrees of freedom of lasso, adaptive lasso and group lasso.
For block canonical regression, we derived a universal rule, an information criterion and an oracle inequality.
The estimator is promising for gravitational wave burst detection and estimation:
we are currently conducting an analysis with physicists on several months of recordings to quantify type I and type II errors,
as well as false discovery rate.
Also, in the spirit of \citet{Park:Hast:l1-r:2007}, generalized linear models could be regularized via smooth James-Stein thresholding.
More generally, SBITE can be employed in other settings than regression  to provide both sparsity and smoothness.

%%%%%%%%%%%%%%%%%%%%%%%%%%%%
\section{Acknowledgements}

I would like to thank C.~Giacobino, L.~Lang and Y.~Velenik for helpful discussions,
and S.~Foffa, R.~Terenzi and the ROG group for providing a sample of the astrophysics data.
Partially supported through Swiss National Science Foundation.

\appendix

\section{Proof of theorem~1}
\label{app:th1}

 Let $s>1$ and $\nu\geq 1$ be fixed. Let $b_j=\|\tilde {\boldsymbol \beta}_j^*\|^{\nu-1}>0$,
 $C_{j,k}=X_j^T X_k$, ${\bf r}_j=X_j\T {\bf Y}-\sum_{k\neq j} C_{j,k} {\boldsymbol \beta}_k$  for all blocks $j=1,\ldots,J$. For the Gaussian likelihood,
(\ref{eq:fixedpointdef}) is given by
${\boldsymbol \beta}_j = \{ 1-\frac{\lambda^\nu}{b_j  \| {\bf r}_j \| } \}_+^s C_{j,j}^{-1}  {\bf r}_j$ for $j=1,\ldots,J$.
Let $F: \real^P \rightarrow \real^P$ defined by $F=(f_1 ,\ldots, f_J)$ with
%\begin{equation}
$f_j({\boldsymbol \beta})={\boldsymbol \beta}_j -  \{ 1-\frac{\lambda^\nu}{b_j  \| {\bf r}_j \| } \}_+^s C_{j,j}^{-1}  {\bf r}_j$,
%\label{eq:fp}
%\end{equation}
where $f_j: \real^P \rightarrow \real^{p_j}$ for $j=1,\ldots,J$.
For $s>1$, $F$ is differentiable on $\real^P$, which is a rectangular region. The fundamental global univalence theorem of \cite{Gale:Nikaido::1965} states
that $F$ is globally univalent on $\real^P$ provided its Jacobian $J({\boldsymbol \beta})$ is a P-matrix (here ``P'' stands for ``positive'')
for every ${\boldsymbol \beta}\in \real^P$. 
A (not necessarily symmetric) real square matrix is a P-matrix if all of its principal minors are positive.
If so, then the ${\bf 0}$-vector in particular has at most one preimage by $F$ and the smooth block iterative thresholding estimate
$\hat {\boldsymbol \beta}^{\rm SBITE}=F^{-1}({\bf 0})$ is unique.

To prove the Jacobian $J({\boldsymbol \beta})$ of $F$ is a P-matrix, let us determine its entries. Clearly $J({\boldsymbol \beta})$ has ones on its diagonal since
${\bf r}_j$ does not depend on ${\boldsymbol \beta}_j$.
For its other entries, consider any point ${\boldsymbol \beta} \in \real^P$, and
 let ${\cal I}_0$ be the set of indices $j$ for which
the inequality $b_j\|{\bf r}_{j}\| \leq \lambda^\nu$ is true;
let $p_0=\sum_{j \in {\cal I}_0} p_j$ and $j_0=|{\cal I}_0|$.
Permuting variables if necessary, the satisfied inequalities are for $j=1, \ldots,j_0$.
Hence, the first $p_0$ lines of the Jacobian at ${\boldsymbol \beta}$ are the $p_0\times P$ matrix $[I_{p_0} \ 0_{p_0 \times (P-p_0)}]$.
For the remaining blocks $j\in \{j_0+1,\ldots,J\}$, the Jacobian is a block matrix with blocks
$$
J_{j,k}=\left \{ \begin{array}{ll} \frac{1}{d_{j}} C_{j,j}^{-1}C_{j,k}, & \ j\neq k \\
                  I_{p_j}= C_{j,j}^{-1}C_{j,j},& j=k
                 \end{array} \right . {\rm for} \ k=j_0+1,\ldots,J,
$$
where $d_{j}=\frac{w_j^s}{1 -s+s w_j}$  with  $1/w_j=1- \frac{\lambda^\nu}{b_j\|{\bf r}_{j}\|}\in(0,1)$.
It is straightforward to show that $1< d_{j}<\infty$ when $s>1$. Hence the Jacobian is
\begin{equation}
J({\boldsymbol \beta})=\left ( \begin{array}{cc} I_{p_0} & 0_{p_0 \times (P-p_0)} \\ B & D^{\bar {\cal I}_0} C^{\bar {\cal I}_0}_{>1} \end{array} \right ),
\label{eq:Jacobian}
\end{equation}
where $B$ is some $(P-p_0)\times p_0$ matrix, $D^{\bar {\cal I}_0}={\rm diag}(C_{j,j}^{-1}/d_j, j=j_0+1,\ldots,J)$, and
%$C^{\bar {\cal I}_0}_{>1}$ is equal to $(X^{\bar {\cal I}_0})\T X^{\bar {\cal I}_0}$ except that the block diagonal elements are multiplied by $d_j>1$, $j=j_0+1,\ldots,J$.
%We can also write $C^{\bar {\cal I}_0}_{>1}=(X^{\bar {\cal I}_0})\T X^{\bar {\cal I}_0}+H$, where $H={\rm diag}((d_j-1) C_{j,j}, j=j_0+1,\ldots,J)$
$C^{\bar {\cal I}_0}_{>1}=(X^{\bar {\cal I}_0})\T X^{\bar {\cal I}_0}+H$.
Since  $H={\rm diag}((d_j-1) C_{j,j}, j=j_0+1,\ldots,J)$ is positive definite if $C_{j,j}=X_j\T X_j>0$ (recall that $d_j>1$ when $s>1$) for all $j=1,\ldots,J$,
then $C^{\bar {\cal I}_0}_{>1}$ is a positive definite matrix when $s>1$.
Consequently, $|J({\boldsymbol \beta})|=|D^{\bar {\cal I}_0}||C^{\bar {\cal I}_0}_{>1}|>0$;
positivity is also verified for all principal minors of $J({\boldsymbol \beta})$ that have the same structure as (\ref{eq:Jacobian}).
So $F$ is an injective function and ${\bf 0}$ has a unique preimage. The Jacobian being invertible, the implicit function theorem guarantees the preimage
is also continuously differentiable with respect to the data.

\section{Proof of theorem~2}
\label{app:th2}

For Gaussian likelihood, the solution to (\ref{eq:fixedpointdef}) is the system of nonlinear equations:
\begin{eqnarray}
  \hat \beta_p &=& \{ 1-\frac{\lambda^\nu}{|\tilde \beta_p^*|^{\nu-1}  |r_p| } \}_+^s   r_p / \| {\bf x}_p \|_2^2 \label{eq:SBITGauss} \\
 & {\rm with} & r_p={\bf x}_p\T {\bf Y}-\sum_{q\neq p} {\bf x}_p^T {\bf x}_q \hat \beta_q , \quad p=1,\ldots,P. \nonumber
\end{eqnarray}
% Using the implicit definition of SBITE (\ref{eq:SBITGauss}) for unit blocks,
% the implicit function theorem states that SBITE is continuously differentiable with respect to the data.
Hence one finds 
\begin{eqnarray}
\frac{\partial \hat \beta_p({\bf Y})}{\partial Y_n}&=&\left \{
\begin{array}{ll}
 0 & {\rm if}\ \hat \beta_p({\bf Y})=0,\\
 \frac{1}{\|{\bf x}_p \|_2^2} ( v_p(x_{np}-{\bf x}_p\T X_{-p} \nabla_n \hat{\boldsymbol \beta}({\bf Y}))+u_p)  & {\rm else,}
\end{array}
\right .
\label{eq:partderpn}
\end{eqnarray}
where
$$
u_p=\frac{s (\nu-1)a_{pn}(w_p-1)r_{p}}{\tilde \beta_p^* w_p^s},\ v_p= \frac{1-s+s w_p}{w_p^s}\quad {\rm and} \quad 1/w_p=1-\frac{\lambda^\nu}{|\tilde \beta_p^*|^{\nu-1}| r_{p}|}.
$$
Let ${\cal I}_0=\{p\in\{1,\ldots,P\}: \hat \beta_p({\bf Y})= 0\}$,
%let $\bar {\cal I}_0$ be its complement,
and let $X^{\bar {\cal I}_0}$ be the columns of $X$ with an index in $\bar {\cal I}_0$.
Rewriting ({\ref{eq:partderpn}),  
 the entries of $\nabla_n \hat{\boldsymbol \beta}( {\bf Y}) =: {\bf h}_n$ are
$$
(\nabla_n \hat{\boldsymbol \beta}( {\bf Y}))_p = \left \{
\begin{array}{ll}
 0 & p \in {\cal I}_0 \\
 h_{n,p} & p \in \bar {\cal I}_0
\end{array}
\right . ,
$$
where ${\bf h}_n^{\bar {\cal I}_0}$ 
are solution to the following system of $|{\bar {\cal I}_0}|$ linear equations:
\begin{equation}
{\bf x}_{p} \T X^{\bar {\cal I}_0} D_{p}^{\bar {\cal I}_0} {\bf h}_n^{\bar {\cal I}_0}=x_{n,p}+\frac{u_p}{v_p} \quad \mbox{for all}\ p\in \bar {\cal I}_0.
\label{eq:lineqgrad}
\end{equation}
Here $D_{p}^{\bar {\cal I}_0}$ is the identity matrix except that its $p$th diagonal element is
$D_{p,p}^{\bar {\cal I}_0} =v_p^{-1}$

%$$
%{\bf x}_p\T X^{\bar {\cal I}_0} D_p^{\bar {\cal I}_0} {\bf h}_n^{\bar {\cal I}_0}=z_{ni}\quad {\rm for} \quad p=p_i, \ i=1,\ldots,{|\bar {\cal I}_0|},
%$$
%where $D_p^{\bar {\cal I}_0}$ is given in (\ref{eq:Dpii})
%and $z_{ni}$ is given in (\ref{eq:zni}).
 %for  $p=p_i \in \bar {\cal I}_0$ and $i=1,\ldots,{|{\cal I}_0|}$.
 The matrix of the linear system (\ref{eq:lineqgrad}) is $(X^{\bar {\cal I}_0})\T X^{\bar {\cal I}_0}$
which diagonal elements are multiplied by $D_{p,ii}^{\bar {\cal I}_0}$.
%$$
%D_{p,ii}^{\bar {\cal I}_0}=\frac{\{\eta^{\rm soft}_{\lambda^\nu}(|\hat \alpha_p^{\rm OLS}|^{\nu-1}|{\bf r}_{-p}^{\rm T} %{\bf x}_p|)\}^{1-s}}{s  |\hat \alpha_p^{\rm OLS}|^{\nu (1-s)}}.
%$$
Moreover  $w_p > 1$, so  all $D_{p,ii}^{\bar {\cal I}_0}> 1$ since $f(w)=w^s/(1-s+s w)$
satisfies $f(1)=1$ and $f'(w)> 0$ for $w> 1$. This guarantees existence of a solution ${\bf h}_n^{\bar {\cal I}_0}$ when $s=1$ if the column of $X$ are linearly independent,
and when $s>1$ otherwise (i.e., $s$ plays the role of a ridge parameter).

\section{Proof of theorem~3}
\label{app:th3}

For $s=1$, each $\hat \rho_n$ is strictly increasing from $\lambda=0$ to $\lambda=|Y_n|$, and is then constant after a jump of size $2\nu$.
So ${\rm TV}^{(s=1)}(\hat \rho_n)= Y_n^2+4\nu-2$ and
${\rm TV}^{(s=1)}({\rm SURE})=\sum_{n=1}^N {\rm TV}^{(s=1)}(\hat \rho_n)$.
For $s>1$, the triangular inequality gives ${\rm TV}^{(s>1)}({\rm SURE})\leq \sum_{n=1}^N {\rm TV}^{(s>1)}(\hat \rho_n)$, and simple calculations lead
to  ${\rm TV}^{(s>1)}(\hat \rho_n)=2 \hat \rho_n((\tilde \lambda_n,\nu,s), \alpha_n)-Y_n^2$,
where $\tilde \lambda_n$ is solution to $\tilde x_n=(1-\tilde \lambda_n^\nu/|Y_n|^\nu)$ with $\tilde x_n\in(0,1)$ the unique root to
$$
%\frac{\partial}{\partial x} \hat \rho_n((x;\nu,s), \alpha_n)= -2 s Y_n^2 (1 - x^s) x^{s-1} + 2 (s-1)x^{s-2}(\nu s(1-x)+x)+2x^{s-1}(-\nu s +1)\equiv 0
\frac{\partial}{\partial x} \hat \rho_n((x;\nu,s), \alpha_n)\propto - s Y_n^2 (1 - x^s) x^{s-1} +  (s-1)x^{s-2}(\nu s(1-x)+x)+x^{s-1}(-\nu s +1)\equiv 0
$$
i.e., $Y_n^2 x (1-x^s)=\nu(s-1)+x (1-\nu s)$. Hence ${\rm TV}^{(s>1)}(\hat \rho_n)=Y_n^2+4\nu \tilde x^{s-1}-2-2Y_n^2 \tilde x^{2s}\leq {\rm TV}^{(s=1)}(\hat \rho_n)$.
Moreover
$\frac{\partial}{\partial Y_n}  {\rm TV}^{(s>1)}(\hat \rho_n) =4 Y_n (1-\tilde x_n^s)^2-2Y_n \leq \frac{\partial}{\partial Y_n}  {\rm TV}^{(s=1)}(\hat \rho_n)=2Y_n$
for ${Y_n \geq 0}$, and
$
\frac{\partial}{\partial \nu}  {\rm TV}^{(s>1)}(\hat \rho_n) =4s \tilde x_n^{s-1}(1-\tilde x_n) \leq 4(1-1/s)^{s-1} \leq \frac{\partial}{\partial \nu}  {\rm TV}^{(s=1)}(\hat \rho_n)=4
$.
At the limit when $Y_n$ tends to zero and for large $\nu$, we have
$\frac{\partial}{\partial \nu}  {\rm TV}^{(s>1)}(\hat \rho_n) \stackrel{\cdot}{=}  4 (1-\frac{1}{s})^{s-1} \geq 4 \exp(-1)$
and the lower bound is reached as $\nu$ grows if for instance $s=2\log \nu+1$.

%%%%%%%%%%%%%%%
\section{Proof of theorem~4}
\label{app:th4}
% 
% The $\ell_2$ risk is defined as 
% \begin{eqnarray*}
% \rho(\lambda,{\boldsymbol \alpha})&=&\sum_{n=1}^N \rho_n(\lambda,{\boldsymbol \alpha}_n)=\sum_{n=1}^N {\rm E} \|\hat {\boldsymbol \alpha}_{n} -  {\boldsymbol \alpha}_n\|_2^2.
% \end{eqnarray*}
% Following \citet{Dono94b}, consider first the oracle predictive performance of the diagonal projection estimator $\hat {\boldsymbol \alpha}_n=\delta {\bf Y}$, where $\delta_n \in \{0,1\}$. The corresponding risk is
% $$
% \rho_n(\delta_n,{\boldsymbol \alpha}_n)=\left \{  \begin{array}{ll}
% | {\boldsymbol \alpha}_n |^2, & {\rm if} \ \delta_n=0 \\
% Q,  & {\rm if} \ \delta_n=1
% \end{array} \right . ,
% $$
% Hence for the oracle hyperparameters $\delta_n^*=1_{\{| {\boldsymbol \alpha}_n|^2>Q\}}$ for $n=1,\ldots,N$, the oracle risk is
% $$
% \rho^*({\boldsymbol \delta},{\boldsymbol \alpha})=\sum_{n=1}^N \min(|{\boldsymbol \alpha}_n |^2,Q).
% $$

The SBIT estimator $(\hat{\boldsymbol \alpha}_n)_{\lambda,\nu;s}$ defined in (\ref{eq:SJS}) with $Q$ fixed has risk
%(1-\frac{\lambda^\nu}{\|{\bf Y}_n\|_2^\nu})_+^s  {\bf Y}_n
\begin{eqnarray}
\rho_n((\lambda,\nu,s),{\boldsymbol \alpha}_n)&=&{\rm E} \| (\hat{\boldsymbol \alpha}_n)_{\lambda,\nu;s}  -  {\boldsymbol \alpha}_n\|_2^2 \nonumber \\
&=& Q + {\rm E} \| (\hat{\boldsymbol \alpha}_n)_{\lambda,\nu;s}  -  {\bf Y}_n\|_2^2 -2Q +
2 {\rm E} ({\bf Y }_n-{\boldsymbol \alpha}_n)\T (\hat{\boldsymbol \alpha}_n)_{\lambda,\nu;s} \nonumber  \\
&=&-Q + {\rm E} \| (\hat{\boldsymbol \alpha}_n)_{\lambda,\nu;s}  -  {\bf Y}_n\|_2^2 + 2 \sum_{q=1}^Q {\rm E}
\frac{\partial (\hat{\boldsymbol \alpha}_n)_{\lambda,\nu;s}}{\partial Y_{n}^{(q)}}, \label{eq:riskn1}
\end{eqnarray}
for all $n=1,\ldots,N$, where we used Stein's lemma  for the last term, and where
$$
\{(\hat{\boldsymbol \alpha}^{(q)}_n)_{{\lambda,\nu;s}}  -  Y_{n}^{(q)}\}^2 = \left \{
\begin{array}{ll}
 (Y^{(q)}_{n})^2 & {\rm if}\ \|{\bf Y}_n\|_2 < \lambda \\
 (Y^{(q)}_{n})^2  \{1-  (1-\frac{\lambda^\nu}{\|{\bf Y}_n\|_2^\nu})^s \}^2 & {\rm if}\ \|{\bf Y}_n\|_2 \geq \lambda
\end{array}
\right .
$$
and $\partial (\hat{\boldsymbol \alpha}_{\lambda,\nu;s})_n / \partial Y_{n}^{(q)}$ is given by (\ref{eq:partialalphaSURE}).
From (\ref{eq:riskn1}) and using the inequality $(1-(1-\epsilon)^s))^2\leq s^2 \epsilon^2$ for $0\leq\epsilon\leq1$ and $s\geq 1$, one gets two inequalities.
First we have
\begin{eqnarray}
 \rho_n((\lambda,\nu,s),{\boldsymbol \alpha}_n)&\leq& -Q + \lambda^2 {\rm P}( \|{\bf Y}_n\|_2 < \lambda) + s^2 \lambda^2 {\rm P}( \|{\bf Y}_n\|_2 > \lambda) 
 +2 (\nu s + Q) {\rm P}( \|{\bf Y}_n\|_2 > \lambda)  \nonumber \\
 & \leq & Q + 2 \nu s + s^2 \lambda^2  \nonumber \\
&\leq &\left \{ \begin{array}{ll}
(Q + 2 \nu s + s^2 \lambda^2)(Q/N+Q) & {\rm if}\ Q\geq 1 \\
(Q + 2 \nu s + s^2 \lambda^2)(Q/N+\|{\boldsymbol \alpha}_n\|_2^2) & {\rm if}\  \|{\boldsymbol \alpha}_n\|_2^2 \geq 1
                \end{array} \right . \label{ineq:1}.
\end{eqnarray}

Second, we show below that
\begin{equation}
\rho_n((\lambda,\nu,s),{\boldsymbol \alpha}_n)\leq (1+Q+ \lambda_{N,Q}^2)(\frac{Q+\nu s}{Q}) (Q/N+\|{\boldsymbol \alpha}_n\|_2^2)\quad {\rm if}\
\|{\boldsymbol \alpha}_n\|_2^2\leq 1
\label{ineq:2}
\end{equation}
for $N$ large enough.
So putting (\ref{ineq:1}) and (\ref{ineq:2}) together, and summing over all $n=1,\ldots,N$ leads to the oracle inequality (\ref{eq:OrIneq}).

To show (\ref{ineq:2}) and complete the proof, note that
\begin{eqnarray}
 \rho_n((\lambda,\nu,s),{\boldsymbol \alpha}_n)&=& {\rm E}  \|{\bf Y}_n\|_2^2 - Q
+ \sum_{q=1}^Q {\rm E}\{(Y_{n}^{(q)})^2 [\{1- (1-\frac{\lambda^\nu}{\|{\bf Y}_n\|_2^\nu})^s\}^2-1] 1(\|{\bf Y}_n\|_2 > \lambda) \} \nonumber \\
&& + 2 \sum_{q=1}^Q {\rm E} \{(1-\frac{\lambda^\nu}{\|{\bf Y}_n\|_2^\nu})^{s-1} (\nu s \lambda^\nu \frac{(Y_{n}^{(q)})^2}{\|{\bf Y}_n\|_2^{\nu+2}} + 1 -\frac{\lambda^\nu}{\|{\bf Y}_n\|_2^\nu} ) 1(\|{\bf Y}_n\|_2 > \lambda)\} \nonumber \\
& \leq & \|{\boldsymbol \alpha}_n \|_2^2  + 2 (\nu s+Q) {\rm P}(\|{\bf Y}_n\|_2 > \lambda)=:\|{\boldsymbol \alpha}_n \|_2^2  + \frac{ \nu s + Q}{Q} g(\mu;\lambda) \label{eq:ineq1}
\end{eqnarray}
with $g(\mu;\lambda)=2Q \{1-\exp(-\mu^2/2) \sum_{j=0}^\infty \frac{(\mu^2/2)^{j}}{j!} \frac{s(j+Q/2,\lambda^2/2)}{\Gamma(j+Q/2)}\}$
since $\|{\bf Y}_n\|_2^2$ is noncentral chi-square with $Q$ degrees of freedom and noncentrality parameter 
$\mu^2=\|{\boldsymbol \alpha}_n\|_2^2<1$.
Considering  even $Q$'s for simplicity, Taylor's expansion gives 
%\begin{eqnarray*}
$g(\mu;\lambda)\leq  g(0;\lambda)+\mu g'(0;\lambda) + \mu^2/2 \sup_{x \in [0,1)} |g''(x;\lambda)|$.
%\end{eqnarray*}
First
\begin{eqnarray*}
g(0;\tilde \lambda_{N,Q})&=&2Q(1- \frac{s(Q/2,\tilde \lambda_{N,Q}^2/2)}{\Gamma(Q/2)})\\
    &=& 2Q\exp(-{\tilde \lambda}_{N,Q}^2/2) \sum_{j=0}^{Q/2-1} \frac{({\tilde \lambda_{N,Q}}^2/2)^{j}}{\Gamma(j+1)}\\
    &\leq& 2Q\exp(- \lambda_{N,Q}^2/2) \sum_{j=0}^{Q/2-1} \frac{(\lambda_{N,Q}^2/2)^{j}}{\Gamma(j+1)}\\
    &=&\frac{2Q}{N} \frac{\Gamma(Q/2)}{(\log N)^{Q/2}}(1+\lambda_{N,Q}^2/2+ \sum_{j=2}^{Q/2-1} \frac{(\lambda_{N,Q}^2/2)^{j}}{\Gamma(j+1)}),
\end{eqnarray*}
where the inequality stems from the fact that  $\tilde \lambda_{N,Q}^2\geq \lambda_{N,Q}^2$  for all $Q\geq 2$ and all $N\geq N_0=\exp(\Gamma(Q/2)^{1/(Q/2-1)})$,
and $\tilde \lambda_{N,Q}^2 \stackrel{\cdot} \sim \lambda_{N,Q}^2$ as $N\rightarrow \infty$.
Then %for $\Gamma(Q/2)/(\log N)^{Q/2}\leq 1$, then 
\begin{eqnarray*}
g(0;\tilde \lambda_{N,Q})&\leq& \frac{Q}{N} (1) [2+\lambda_{N,Q}^2 +2\sum_{j=2}^{Q/2-1} e(j,Q,N)]\\
&\leq& \frac{Q}{N} [2+\lambda_{N,Q}^2+2(Q/2-2)(1)]\leq\frac{Q}{N} [Q+\lambda_{N,Q}^2],
\end{eqnarray*}
since $\frac{\Gamma(Q/2)}{(\log N)^{Q/2}}\leq 1$ and $e(j,Q,N)=(1+\frac{Q/2\log \log N-\log \Gamma(Q/2)}{\log N})^j\frac{ \Gamma(Q/2)}{(\log N)^{Q/2-j}\Gamma(j+1)}\leq 1$
 for $N$ large enough. Second, note that
\begin{eqnarray*}
g'(x;\lambda)%&=&2Q x \exp(-x^2/2) (   \sum_{j=0}^\infty \frac{(x^2/2)^{j}}{j!} c_{j,Q}(\lambda^2)  \\ && - \sum_{j=1}^\infty \frac{(x^2/2)^{j-1}}{(j-1)!} [c_{j-1,Q}(\lambda^2)-\frac{(\lambda^2/2)^{j-1+Q/2}\exp(-\lambda^2/2)}{\Gamma(j-1+Q/2+1)}] ) \\
&=& 2Q x \exp(-x^2/2) \exp(-\lambda^2/2)  \sum_{j=0}^\infty \frac{(x^2/2)^{j}}{\Gamma(j+1)} \frac{(\lambda^2/2)^{j+Q/2}}{\Gamma(j+Q/2+1)}] 
\end{eqnarray*}
so $g'(0;\lambda)=0$. Finally
\begin{eqnarray*}
g''(x;\lambda)&=&2Q\exp(-\lambda^2/2)\exp(-x^2/2) (1-x^2) \sum_{j=0}^\infty \frac{(x^2/2)^j}{\Gamma(j+1)} \frac{(\lambda^2/2)^{j+Q/2}}{\Gamma(j+Q/2+1)}\\
&& + 2Q\exp(-\lambda^2/2)\exp(-x^2/2) x^2 \sum_{j=0}^\infty \frac{(x^2/2)^j}{\Gamma(j+1)} \frac{(\lambda^2/2)^{j+Q/2+1}}{\Gamma(j+Q/2+2)}\\
&\leq& 2Q+ 2Q x^2 S (\frac{\lambda^2/2}{Q/2}-1) \leq 2Q+2\lambda^2
\end{eqnarray*}
with $S=\exp(-\lambda^2/2)\exp(-x^2/2)  \sum_{j=0}^\infty \frac{(x^2/2)^j}{\Gamma(j+1)} \frac{(\lambda^2/2)^{j+Q/2}}{\Gamma((j+Q/2+1)}\leq 1$.
The same inequality holds for $-g''(x;\lambda)$. Consequently for $N$ larger than $N_0$, we have
%\begin{equation}
 $g(\mu,\lambda)\leq Q/N (Q+ \lambda_{N,Q}^2) + \mu^2/2 ( 2Q+2 \lambda_{N,Q}^2)$
%\label{eq:gmuleq}
%\end{equation}
for $\mu=\|{\boldsymbol \alpha}_n\|_2<1$.

\bibliographystyle{chicago}
\bibliography{article}

\begin{thebibliography}{}

\bibitem[\protect\citeauthoryear{Antoniadis and Fan}{Antoniadis and
  Fan}{2001}]{AntoFan01}
Antoniadis, A. and Fan, J. (2001).
\newblock Regularization of wavelet approximations (with discussion).
\newblock {\em Journal of the American Statistical Association\/}~{\bf 96},
  939--967.

\bibitem[\protect\citeauthoryear{Bakin}{Bakin}{1999}]{bakin99}
Bakin, S. (1999).
\newblock {\em Adaptive regression and model selection in data mining
  problems}.
\newblock Ph.\ D. thesis, Australian National University, Canberra.

\bibitem[\protect\citeauthoryear{Bertsekas}{Bertsekas}{1999}]{Bert}
Bertsekas, D.~P. (1999).
\newblock {\em Nonlinear Programming}.
\newblock Belmont, MA: Athena Scientific.

\bibitem[\protect\citeauthoryear{Breiman}{Breiman}{1995}]{Brei:bett:1995}
Breiman, L. (1995).
\newblock Better subset regression using the nonnegative garrote.
\newblock {\em Technometrics\/}~{\bf 37}, 373--384.

\bibitem[\protect\citeauthoryear{Breiman}{Breiman}{1996}]{B96}
Breiman, L. (1996).
\newblock Heuristics of instability and stabilization in model selection.
\newblock {\em Annals of Statistics\/}~{\bf 24}, 2350--2383.

\bibitem[\protect\citeauthoryear{Cai}{Cai}{1999}]{Cai:adap:1999}
Cai, T.~T. (1999).
\newblock Adaptive wavelet estimation: a block thresholding and oracle
  inequality approach.
\newblock {\em The Annals of Statistics\/}~{\em 27\/}(3), 898--924.

\bibitem[\protect\citeauthoryear{Cand\`es}{Cand\`es}{2005}]{Candes:2005}
Cand\`es, E. (2005).
\newblock Modern statistical estimation via oracle inequalities.
\newblock {\em Acta Numerica\/}~{\bf 15}, 257--325.

\bibitem[\protect\citeauthoryear{Chen, Donoho, and Saunders}{Chen
  et~al.}{1999}]{CDS99}
Chen, S.~S., Donoho, D.~L., and Saunders, M.~A. (1999).
\newblock Atomic decomposition by basis pursuit.
\newblock {\em SIAM Journal on Scientific Computing\/}~{\bf 20(1)}, 33--61.

\bibitem[\protect\citeauthoryear{Daubechies, Defrise, and Mol}{Daubechies
  et~al.}{2004}]{DaubInvProb04}
Daubechies, I., Defrise, M., and Mol, C.~D. (2004).
\newblock A iterative thresholding algorithm for linear inverse problems with a
  sparsity constraint.
\newblock {\em Communications on Pure and Applied Mathematics\/}~{\bf 57},
  1413--1457.

\bibitem[\protect\citeauthoryear{Donoho and Johnstone}{Donoho and
  Johnstone}{1994}]{Dono94b}
Donoho, D.~L. and Johnstone, I.~M. (1994).
\newblock Ideal spatial adaptation via wavelet shrinkage.
\newblock {\em Biometrika\/}~{\bf 81}, 425--455.

\bibitem[\protect\citeauthoryear{Donoho and Johnstone}{Donoho and
  Johnstone}{1995}]{Dono95i}
Donoho, D.~L. and Johnstone, I.~M. (1995).
\newblock Adapting to unknown smoothness via wavelet shrinkage.
\newblock {\em Journal of the American Statistical Association\/}~{\bf 90},
  1200--1224.

\bibitem[\protect\citeauthoryear{Efron, Hastie, Johnstone, and
  Tibshirani}{Efron et~al.}{2004}]{Efro:Hast:John:Tibs:leas:2004}
Efron, B., Hastie, T., Johnstone, I., and Tibshirani, R. (2004).
\newblock Least angle regression.
\newblock {\em The Annals of Statistics\/}~{\em 32\/}(2), 407--499.

\bibitem[\protect\citeauthoryear{Embrechts, Kluppelberg, and Mikosch}{Embrechts
  et~al.}{1997}]{Embr:Klup:Miko:mode:1997}
Embrechts, P., Kluppelberg, C., and Mikosch, T. (1997).
\newblock {\em Modelling Extremal Events: For Insurance and Finance}.
\newblock Springer-Verlag Inc.

\bibitem[\protect\citeauthoryear{Fan and Li}{Fan and
  Li}{2001a}]{Fan:Li:vari:2001}
Fan, J. and Li, R. (2001a).
\newblock Variable selection via nonconcave penalized likelihood and its oracle
  properties.
\newblock {\em Journal of the American Statistical Association\/}~{\em
  96\/}(456), 1348--1360.

\bibitem[\protect\citeauthoryear{Fan and Li}{Fan and Li}{2001b}]{FanLi01}
Fan, J. and Li, R. (2001b).
\newblock Variable {S}election via {N}onconcave {P}enalized {L}ikelihood and
  {I}ts {O}racle {P}ropoerties.
\newblock {\em Journal of the American Statistical Association\/}~{\bf 96},
  1348--1360.

\bibitem[\protect\citeauthoryear{Fu}{Fu}{1998}]{Fu:1998}
Fu, W.~J. (1998).
\newblock Penalized regressions: {T}he bridge versus the lasso.
\newblock {\em Journal of Computational and Graphical Statistics\/}~{\bf 7},
  397--416.

\bibitem[\protect\citeauthoryear{Gale and Nikaido}{Gale and
  Nikaido}{1965}]{Gale:Nikaido::1965}
Gale, D. and Nikaido, Y. (1965).
\newblock The {J}acobian matrix and global univalence of mappings.
\newblock {\em Mathematischen Annalen\/}~{\bf 159}, 81--93.

\bibitem[\protect\citeauthoryear{Golub, Heath, and Wahba}{Golub
  et~al.}{1979}]{Golu:Heat:Wahb:gene:1979}
Golub, G.~H., Heath, M., and Wahba, G. (1979).
\newblock Generalized cross-validation as a method for choosing a good ridge
  parameter.
\newblock {\em Technometrics\/}~{\bf 21}, 215--223.

\bibitem[\protect\citeauthoryear{Hoerl and Kennard}{Hoerl and
  Kennard}{1970}]{ridgeHK}
Hoerl, A.~E. and Kennard, R.~W. (1970).
\newblock Ridge regression: biased estimation for nonorthogonal problems.
\newblock {\em Technometrics\/}~{\bf 12}, 55--67.

\bibitem[\protect\citeauthoryear{James and Stein}{James and
  Stein}{1961}]{Jame:Stei:esti:1961}
James, W. and Stein, C. (1961).
\newblock Estimation with quadratic loss.
\newblock In J.~Neyman (Ed.), {\em Proceedings of the Fourth Berkeley Symposium
  on Mathematical Statistics and Probability, Volume 1}, pp.\  361--379.
  University of California Press.

\bibitem[\protect\citeauthoryear{Johnstone and Silverman}{Johnstone and
  Silverman}{2004}]{JS04}
Johnstone, I.~M. and Silverman, B. (2004).
\newblock Needles and straw in haystacks: {E}mpirical {B}ayes estimates of
  possibly sparse sequences.
\newblock {\em Annals of Statistics\/}~{\bf 32}, 1594--1649.

\bibitem[\protect\citeauthoryear{Johnstone and Silverman}{Johnstone and
  Silverman}{2005}]{JS05}
Johnstone, I.~M. and Silverman, B. (2005).
\newblock Empirical {B}ayes selection of wavelet thresholds.
\newblock {\em Annals of Statistics\/}~{\bf 33}, 1700--1752.

\bibitem[\protect\citeauthoryear{Johnstone and Silverman}{Johnstone and
  Silverman}{1997}]{John:Silv:wave:1997}
Johnstone, I.~M. and Silverman, B.~W. (1997).
\newblock Wavelet threshold estimators for data with correlated noise.
\newblock {\em Journal of the Royal Statistical Society, Series B:
  Methodological\/}~{\bf 59}, 319--351.

\bibitem[\protect\citeauthoryear{Klimenko and Mitselmakher}{Klimenko and
  Mitselmakher}{2004}]{gravitwave04}
Klimenko, S. and Mitselmakher, G. (2004).
\newblock A wavelet method for detection of gravitational wave bursts.
\newblock {\em Classical and Quantum Gravity\/}~{\bf 21}, 1819--1830.

\bibitem[\protect\citeauthoryear{Nelder and Wedderburn}{Nelder and
  Wedderburn}{1972}]{NW72}
Nelder, J.~A. and Wedderburn, R.~W.~M. (1972).
\newblock Generalized linear models.
\newblock {\em Journal of the Royal Statistical Society, Series A\/}~{\bf 135},
  370--384.

\bibitem[\protect\citeauthoryear{Park and Hastie}{Park and
  Hastie}{2007}]{Park:Hast:l1-r:2007}
Park, M.~Y. and Hastie, T. (2007).
\newblock L1-regularization path algorithm for generalized linear models.
\newblock {\em Journal of the Royal Statistical Society, Series B: Statistical
  Methodology\/}~{\em 69\/}(4), 659--677.

\bibitem[\protect\citeauthoryear{Percival}{Percival}{1995}]{Perc:on:1995}
Percival, D.~P. (1995).
\newblock On estimation of the wavelet variance.
\newblock {\em Biometrika\/}~{\bf 82}, 619--631.

\bibitem[\protect\citeauthoryear{Sardy}{Sardy}{2000}]{minimaxC:2000}
Sardy, S. (2000).
\newblock Minimax threshold for denoising complex signals with {W}aveshrink.
\newblock {\em {IEEE} {T}ransactions on {S}ignal {P}rocessing\/}~{\em 48\/}(4),
  1023--1028.

\bibitem[\protect\citeauthoryear{Sardy}{Sardy}{2008}]{SardyISI08}
Sardy, S. (2008).
\newblock On the practice of rescaling covariates.
\newblock {\em International Statistical Review\/}~{\bf 76}, 285--297.

\bibitem[\protect\citeauthoryear{Sardy}{Sardy}{2009}]{SardySLIC09}
Sardy, S. (2009).
\newblock Adaptive posterior mode estimation of a sparse sequence for model
  selection.
\newblock {\em Scandinavian Journal of Statistics\/}~{\bf 36}, 577--601.

\bibitem[\protect\citeauthoryear{Sardy, Bruce, and Tseng}{Sardy
  et~al.}{2000}]{sardyJCGS}
Sardy, S., Bruce, A.~G., and Tseng, P. (2000).
\newblock Block coordinate relaxation methods for nonparametric wavelet
  denoising.
\newblock {\em Journal of Computational and Graphical Statistics\/}~{\bf 9},
  361--379.

\bibitem[\protect\citeauthoryear{Sardy and Tseng}{Sardy and
  Tseng}{2004}]{SardyTseng04}
Sardy, S. and Tseng, P. (2004).
\newblock On the statistical analysis of smoothing by maximizing dirty markov
  random field posterior distributions.
\newblock {\em Journal of the American Statistical Association\/}~{\bf 99},
  191--204.

\bibitem[\protect\citeauthoryear{Serroukh, Walden, and Percival}{Serroukh
  et~al.}{2000}]{Serr:Wald:Perc:stat:2000}
Serroukh, A., Walden, A.~T., and Percival, D.~B. (2000).
\newblock Statistical properties and uses of the wavelet variance estimator for
  the scale analysis of time series.
\newblock {\em Journal of the American Statistical Association\/}~{\em
  95\/}(449), 184--196.

\bibitem[\protect\citeauthoryear{Stein}{Stein}{1981}]{Stein:1981}
Stein, C. (1981).
\newblock Estimation of the {M}ean of a {M}ultivariate {N}ormal {D}istribution.
\newblock {\em The Annals of Statistics\/}~{\bf 9}, 1135--1151.

\bibitem[\protect\citeauthoryear{Tibshirani}{Tibshirani}{1996}]{Tibs:regr:1996}
Tibshirani, R. (1996).
\newblock Regression shrinkage and selection via the lasso.
\newblock {\em Journal of the Royal Statistical Society, Series B:
  Methodological\/}~{\bf 58}, 267--288.

\bibitem[\protect\citeauthoryear{Tibshirani, Saunders, Rosset, Zhu, and
  Knight}{Tibshirani et~al.}{2005}]{Tibs:Saun:Ross:Zhu:Knig:spar:2005}
Tibshirani, R., Saunders, M., Rosset, S., Zhu, J., and Knight, K. (2005).
\newblock Sparsity and smoothness via the fused lasso.
\newblock {\em Journal of the Royal Statistical Society, Series B: Statistical
  Methodology\/}~{\em 67\/}(1), 91--108.

\bibitem[\protect\citeauthoryear{Wahba}{Wahba}{1990}]{Wahb:spli:1990}
Wahba, G. (1990).
\newblock {\em Spline Models for Observational Data}.
\newblock SIAM.

\bibitem[\protect\citeauthoryear{Yuan and Lin}{Yuan and
  Lin}{2006}]{Yuan:Lin:mode:2006}
Yuan, M. and Lin, Y. (2006).
\newblock Model selection and estimation in regression with grouped variables.
\newblock {\em Journal of the Royal Statistical Society, Series B: Statistical
  Methodology\/}~{\em 68\/}(1), 49--67.

\bibitem[\protect\citeauthoryear{Zou}{Zou}{2006}]{Zou:adap:2006}
Zou, H. (2006).
\newblock The adaptive {LASSO} and its oracle properties.
\newblock {\em Journal of the American Statistical Association\/}~{\bf 101},
  1418--1429.

\bibitem[\protect\citeauthoryear{Zou and Hastie}{Zou and
  Hastie}{2005}]{ZouHastie05}
Zou, H. and Hastie, T. (2005).
\newblock Regularization and variable selection via the elastic net.
\newblock {\em Journal of the Royal Statistical Society, Series B\/}~{\bf 67},
  301--320.

\bibitem[\protect\citeauthoryear{Zou, Hastie, and Tibshirani}{Zou
  et~al.}{2007}]{ZHT07}
Zou, H., Hastie, T., and Tibshirani, R. (2007).
\newblock On the "degrees of freedom" of the lasso.
\newblock {\em The Annals of Statistics\/}~{\bf 35}, 2173--2192.

\end{thebibliography}

% \noindent
% Sylvain Sardy, Department of Mathematics, University of Geneva, 2-4 rue du Li\`evre, Case postale 64, 1211 Gen\`eve 4, Switzerland. \\
% E-mail: Sylvain.Sardy@unige.ch

\end{document}